\documentclass[%
 reprint,
 amsmath,amssymb,
 aps,
]{revtex4-2}

\usepackage{graphicx}% Include figure files
\usepackage{dcolumn}% Align table columns on decimal point
\usepackage{bm}% bold math
\usepackage{amsthm}
\newtheorem{theorem}{Theorem}

\begin{document}

\title{The Completeness of Eigenstates in Quantum Mechanics}

\author{Guoping Zhang}
\affiliation{
 Shenzhen Institute for Quantum Science and Engineering, Southern University of Science and Technology\\
Shenzhen 518055,11930047@mail.sustech.edu.cn}

\date{\today}

\begin{abstract}

\begin{description}
\item[Abstract]
We delineate the scope of research on the completeness of eigenstates in quantum mechanics. Based on the limit of the potential function at infinity, the proof of completeness is divided into eight cases, and theoretical proofs or numerical simulations are provided for each case. We present the definition of orthonormalization for general free states and the solution to the normalization coefficients, as well as a general set of initial states, which simplifies and concretizes the proof of completeness. Additionally, we define the spectral function for continuous energy eigenvalues. By taking the spectral function as the original integral variable of the expansion function, the relationship between the measured probability amplitude and the expansion function is endowed with the physical meaning of coordinate-momentum transformation.
\end{description}

\end{abstract}

\maketitle

\section{\label{sec:level00}introduction}

The Schrödinger equation (SE) is the fundamental dynamical equation describing quantum systems, and its linearity directly gives rise to the superposition principle. The completeness of a set of orthonormal eigenstates precisely characterizes this superposition principle. Completeness requires that any state of the system can be expressed as a linear combination of these eigenstates; conversely, any linear combination of the eigenstates also constitutes a possible state of the system \cite{Dirac}. This property has been intuitively verified in numerous Young's double-slit experiments \cite{Young} and enables the prediction of quantum jump probabilities \cite{jump}.Born's probability interpretation \cite{Born} further endows eigenstates with physical significance: the squared modulus of the coefficient of each eigenstate in a superposition corresponds to the probability of the particle being in that state. The physical meaning of completeness is interpreted as a representation transformation \cite{Transformation}.

The time-independent SE is a type of (generalized) Sturm-Liouville (S-L) equation. Mathematical research on the completeness of solutions to S-L equations dates back to the Hilbert-Schmidt theory in 1907 \cite{Hilbert}. In 1910, Weyl extended S-L equations from finite to infinite intervals. Courant and Hilbert proved the completeness of the discrete spectrum (bound states) in 1924 \cite{discrete}. Titchmarsh systematically developed the eigenfunction expansion theory for singular S-L operators in 1946, establishing the completeness of the continuous spectrum (free states) and mixed spectrum (coexisting bound and free states) under certain conditions \cite{Titchmarsh}. Kodaira comprehensively summarized previous results in 1949 and developed simplified proof methods for completeness \cite{Kodaira}\cite{Well}. Since then, progress in completeness research has been slow—no general proof for the completeness of S-L equations has been achieved to date. Additionally, mathematical proofs of completeness are complex and lack clear physical interpretation.

For quantum mechanics, however, it suffices to prove the completeness of SE solutions. Furthermore, the potential function can be restricted to forms encountered in physical scenarios, eliminating the need for generalized discussions. We first classify potential functions into two categories: symmetric and asymmetric potentials (Sec. III and IV). Following Titchmarsh's classification method \cite{Titchmarsh1}, we further divide these two categories into 8 types based on four possible asymptotic behaviors of the potential at infinity (infinite, constant, finite oscillatory, and infinite oscillatory). Among these 8 types, only one has an existing general theoretical proof (Subsection IIIA). For the remaining cases, we provide theoretical proofs for some and conduct completeness tests via numerical simulations for others. In the course of these proofs, the spectral function retains physical meaning (momentum spectrum), and the form of completeness aligns with the interpretation of representation transformation (coordinate and momentum representations).

\section{\label{sec:level01}common issues}

\subsection{\label{sec:level011}A set of general initial states}
The initial state serves as the test object for completeness. Once the set of general initial states is determined, it becomes convenient to judge the completeness of the set of eigenstates.

\paragraph{Property of the complete set}
The set of eigenstates of an infinite square well defined in \( x \in [-{\sigma}/2, {\sigma}/2] \) is complete for functions whose primitive functions are piecewise continuous in \( (-{\sigma}/2, {\sigma}/2) \) and have square-integrable first derivatives in their continuous regions\cite{discrete}.

\paragraph{Localized state}
If the space of the initial state is localized, one can always choose the region \([-{\sigma}/2, {\sigma}/2]\) to contain it.

\paragraph{Nonlocalized state}
If the space of the initial state is nonlocal, let ${\sigma}$ gradually increase to \( \infty \). As ${\sigma}$ increases, every value of ${\sigma}$ satisfies conclusion $b$. In this sense, we argue that nonlocal initial states can be expanded in terms of the set of eigenstates of an infinite square well.

Based on the three conclusions above, we conclude that the set of eigenstates of the infinite square well can serve as a set of generic initial states. 
A set of generic initial states can be set as  
\begin{equation}
{\Psi }_{j}(x) = 
\begin{cases} 
\sqrt{\dfrac{2}{\sigma}} \sin\left( \dfrac{j \pi}{\sigma} (x - \tau) \right), &\tau < x < \tau + \sigma \\
0 ,&\text{otherwise}
\end{cases}.
\label{eq:initial states}
\end{equation}
where \( j = 1, 2, 3 \ldots \), and \( \tau, \sigma \) are arbitrary given real numbers.
Now, we can determine whether a given set of eigenstates is complete. 

\begin{theorem}
If a set of eigenstates is complete for the wavefunction ${\Psi }_{j}(x)$ in Eq. (1), then it is complete.
\end{theorem}
The time evolution of the initial state is equivalent to the temporal evolution of the system following the instantaneous removal of the left and right boundaries of the box confining the potential.

\subsection{\label{sec:level2}Orthonormalization of free states}
In the remainder of the paper, the term "\underline{side}" refers to the infinity of one-dimensional space. The "orthonormalization" method for free states:  
\[\left \langle {\psi }_{\kappa }|{\psi }_{\kappa '}\right \rangle=\int_{-\infty}^{+\infty} \frac{1}{2\pi} \exp\left[i(\kappa - \kappa')x\right] dx\]
$= \int_{-\infty}^{x_1} \frac{1}{\pi} \exp\left[i(\kappa - \kappa')x\right] dx
= \int_{x_2}^{+\infty} \frac{1}{\pi} \exp\left[i(\kappa - \kappa')x\right] dx $
\begin{equation}
= I(\kappa - \kappa') = \begin{cases} 
\text{finite large number}, & \text{if } \kappa \neq \kappa' \\
\infty, & \text{if } \kappa = \kappa'
\end{cases}.
\label{eq:orthonormalization}
\end{equation}
Where $x_1,x_2$ are arbitrary finite coordinates and $ \kappa(\varepsilon ) , \kappa'(\varepsilon) $ are {spectral function($\varepsilon$ is the continuous energy eigenvalue,proceed to the next subsection). Here, for "orthonormalization", instead of $\delta(\kappa - \kappa')$, we use the symbol $I(\kappa - \kappa')$. this is because the former is already employed in the context of completeness. For Eq. (2), one can understand it as follows: the average probability amplitude of the free state on both sides for "normalization" is \( \sqrt{1/(2\pi)} \), while the average probability amplitude of the free state restricted to one side only for "normalization" is \( \sqrt{1/\pi} \). Such Orthonormalization is derived from the ratio of infinities (or equivalently, Fourier transforms and sine-cosine transforms \cite{Cos-sin}). Thus, the actual meaning of "=" here is equivalent infinity symbol "$\sim$".

Let the energy eigenstates be ${\phi }_{\kappa \varepsilon }\left ( x\right )$ and the orthonormalized energy eigenstates be ${\psi }_{\kappa \varepsilon }\left ( x\right )=A\left ( \kappa ,\varepsilon \right ){\phi }_{\kappa \varepsilon }\left ( x\right )$. From Eq. 2, the normalization coefficient is given by (without loss of generality, we take $A\left ( \kappa ,\varepsilon \right )$ as a positive real number): 
\begin{equation}
{A(\kappa ,\varepsilon)}^2={\lim_{a\rightarrow +\infty }\frac{\int_{-a}^{+a}{{{\left |{\frac{1}{\sqrt{2\pi }}}\right |}^2}dx}}{\int_{-a}^{+a}{{{\left |{{\phi_{\kappa \varepsilon}}(x)}\right |}^2}dx}}},
\label{eq:normalization coefficient 1}
\end{equation}
or
\begin{equation}
{A(\kappa ,\varepsilon)}^2={\lim_{a\rightarrow \infty }\frac{\int_{x_0}^{a}{{{\left |{\frac{1}{\sqrt{\pi }}}\right |}^2}dx}}{\int_{x_0}^{a}{{{\left |{{\phi_{\kappa \varepsilon}}(x)}\right |}^2}dx}}}.
\label{eq:normalization coefficient 2}
\end{equation}
Accounting for spatial symmetry and to avoid the effects of the order of limits, we adopt the Cauchy principal value integral \cite{Cauchy}.

\subsection{\label{sec:level2}The spectral function}

From Eq. (2), we find that the spectral function \(\kappa(\varepsilon ) \) possesses a well-defined wave number dimension. When we compute the expansion function of the initial state, this operation carries the physical meaning of transforming from the coordinate representation to the momentum representation.

To fully determine the energy eigenstates of a one-dimensional system, one generally requires 1 or 2 quantum numbers. Suppose the system has quantum numbers \(\{\varepsilon\}\) or \(\{\varepsilon, \kappa\}\). For free states, the complete quantum numbers are \(\{\varepsilon, \kappa\}\). Thus, in this manuscript, $\kappa(\varepsilon)$ denotes the spectral function for continuous energy eigenvalues.

In general, $\kappa$ is independent of the potential at finite distances (Eqs. (5), (6)). For measurement normalization and the expansion of the initial state, $\kappa$ serves as the original integration variable (Eqs. (36), (22)). In different systems, \(\kappa\) may take different forms, with the common ones being the wave number (Eq. (6)) and the Bloch wave number \cite{Bloch}.

\section{\label{sec:level1}Part One: Symmetric potential}

\subsection{\label{sec:level11}Infinite potential on two sides}
In a potential of this form, the energy eigenstates are bound states, and their completeness has been resolved \cite{Hilbert}. For the completeness of solutions to the Sturm - Liouville equation, the sufficient condition the potential must satisfy is that there are finitely many discontinuities of finite magnitude in the potential. For continuous states, one can give an example that does not require this condition. The Dirac comb (with infinitely many discontinuities of infinite magnitude, Subsec. D) nevertheless has a set of complete energy eigenstates.

In physics, we also frequently encounter unbound states, such as free states and periodic states (see Secs. B, C, and D), which are continuous states. Continuous states require infinite space---an approximate physical model adopted for the convenience of problem-solving. A brief personal perspective on infinite space: If a free particle or wave could avoid interacting or interfering with itself, the free state (infinite space) may serve as a sound description of "freedom". Perhaps our universe is exactly this way.

\subsection{\label{sec:level2}Equal constant potential on two sides}
In such a potential, the spectrum is continuous or mixed. The completeness of sets containing only pure continuous spectrum has been rigorously established via Fourier transforms (Dirichlet's theorem \cite{Dirichlet}). The universal SFs(let \( m  = \hbar = 1 \)):
\begin{equation}
   {\kappa}^{\pm}(\varepsilon)=\lim_{x\to\pm \infty}\sqrt{2[\varepsilon-V(x)]} 
\label{eq:SFs}
\end{equation}
Here,$\varepsilon$ denotes the energy eigenvalue of the system. If ${\kappa }^{+}(\varepsilon)={\kappa }^{-}(\varepsilon)$, we can conclude that a unique SF :
\begin{equation}
   {\kappa}(\varepsilon)=\lim_{x\to\infty}\sqrt{2[\varepsilon-V(x)]} 
\label{eq:SF}
\end{equation}
There exist many examples of such SFs, and some of them have been proven. Examples proven for \( {\kappa}(\varepsilon)=\sqrt{2\varepsilon } \) include reflectionless potentials \cite{Reflectionless}, \(\delta \) potentials \cite{delta}, and Coulomb potentials \cite{Coulomb}, among others.

We present here an example of a double-well potential (symmetric on two sides but locally asymmetric), whose eigenstates are a mixture of bound and free states.  Numerical simulations show that its ${\kappa}(\varepsilon)$ and completeness are independent of finite potentials at finite distances.We adopt natural units where \(\hbar = m = a =  1 \), a double-well potential
\begin{equation}
    V(x)=\begin{cases}
0,& x<0\\ 
-{V}_{0},& 0<x<1\\
-{V}_{1},& 1<x<2\\
0,& x>2
\end{cases}.
\label{pt:1}
\end{equation}
where ${V}_{0}>{V}_{1}.$ From Eq. \ref{eq:SF} and 7, we find that its spectral function is ${\kappa}(\varepsilon)= \sqrt{2\varepsilon},\varepsilon>0$.

\subsubsection{\label{sec:level21}The set of eigenstates}

\paragraph{Bound states 1}
For $-{V}_{0}<\varepsilon <-{V}_{1} $,
SE :
\[x<0 , x>2,{\psi }''(x)-{k}^{2}\psi(x) =0,{k} =\sqrt{-2\varepsilon_n };\]
\[0<x<1,{\psi }''(x)+{{k}_{1} }^{2}\psi(x) =0,{k}_{1} =\sqrt{2(\varepsilon_n +{V}_{0})};\]
\[1<x<2,{\psi }''(x)-{{k}_{2} }^{2}\psi(x) =0,{k}_{2} =\sqrt{-2(\varepsilon_n +{V}_{1})}.\]
The discrete energy levels $\varepsilon_n$ are determined by the energy level equation below.
\begin{eqnarray}
&&2 k k_1 k_2+k_2\left(k^2-k_1^2\right) \tan{k_1}\nonumber\\
&&+\left(k_1 \left(k^2+k_2^2\right) +k \left(k_2^2-k_1^2\right)\tan{k_1}\right)\tanh{k_2}=0.
\end{eqnarray}
From the continuity of ${\psi}$ and ${\psi}'$ at the boundary, we obtain the bound state equation as:
\begin{equation}
{\psi }_{n}^{1}(x)=A \begin{cases}
e^{k x},& x<0\\ 
{(k/ {k_1}) \sin \left(k_1 x\right)}+\cos \left(k_1 x\right),& 0<x<1\\
B e^{k_2 (x-1)}+C e^{-k_2 (x-1)},& 1<x<2\\
 D e^{-k (x-2)},& x>2
\end{cases}.
 \label{eq:eigenstates21}
\end{equation}
\begin{scriptsize} %  \footnotesize, \scriptsize, \tiny
\[B=\left(\left(k k_2-k_1^2\right) \sin k_1+k_1 \left(k+k_2\right) \cos k_1\right)/\left(2 k_1 k_2\right),\]
\[C=\left(\left(k_1^2+k k_2\right) \sin k_1+k_1 \left(k_2-k\right) \cos k_1\right)/(2 k_1 k_2),\]
\[D=\sinh k_2\left(k \cos k_1-k_1 \sin k_1\right)/{k_2}+\cosh k_2\left(k \sin k_1+k_1 \cos k_1\right)/k_1.\]
\end{scriptsize}
The positive real number $A$ is determined by normalization.

\paragraph{Bound states 2}
For $-{V}_{1}<\varepsilon <0 $,
SE :
\[x<0 , x>2,{\psi }''(x)-{k}^{2}\psi(x) =0,{k} =\sqrt{-2\varepsilon_n };\]
\[0<x<1,{\psi }''(x)+{{k}_{3} }^{2}\psi(x) =0,{k}_{3} =\sqrt{2(\varepsilon_n +{V}_{0})};\]
\[1<x<2,{\psi }''(x)+{{k}_{4} }^{2}\psi(x) =0,{k}_{4} =\sqrt{2(\varepsilon_n +{V}_{1})}.\]
The discrete energy levels $\varepsilon_n$ are determined by the energy level equation below.
\begin{eqnarray}
&&2 k k_3 k_4+k_4\left(k^2-k_3^2\right)\tan{k_3}\nonumber\\
&&+\left(k_3 \left(k^2-k_4^2\right)-k \left(k_3^2+k_4^2\right) \tan{k_3}\right) \tan{k_4}=0.
\end{eqnarray}
The bound states are given by:
\begin{equation}
{\psi }_{n}^{2}(x)=A' \begin{cases}
e^{k x},& x<0\\ 
{(k/ {k_3}) \sin \left(k_3 x\right)}+\cos \left(k_3 x\right),& 0<x<1\\
B' \cos \left(k_4 (x-1)\right)+\\
C' \sin \left(k_4 (x-1)\right),& 1<x<2\\
D' e^{-k (x-2)},& x>2
\end{cases}.
 \label{eq:eigenstates22}
\end{equation}
\begin{scriptsize} %  \footnotesize, \scriptsize, \tiny
\[B'=\left(k \sin{k_3}+k_3 \cos{k_3}\right)/{k_3},C'=\left(k \cos{k_3}-k_3 \sin {k_3}\right)/{k_4},\]
\[D'=\left(k \sin{k_3}+k_3 \cos{k_3}\right)\cos{k_4}/{k_3}+\sin {k_4}\left(k \cos{k_3}-k_3 \sin{k_3}\right)/{k_4}.\] 
\end{scriptsize}
The positive real number $A'$ is determined by normalization.

\paragraph{Free states 1} For $\varepsilon>0$,the particle is incident from the left, with SE:
\[x<0 , x>2,{\psi }''(x)+{\kappa}^{2}\psi(x) =0,{\kappa} =\sqrt{2\varepsilon };\]
\[0<x<1,{\psi }''(x)+{{k}_{5} }^{2}\psi(x) =0,{k}_{5} =\sqrt{2(\varepsilon +{V}_{0})};\]
\[1<x<2,{\psi }''(x)+{{k}_{6} }^{2}\psi(x) =0,{k}_{6} =\sqrt{2(\varepsilon +{V}_{1})}.\]
The free states are given by:
\begin{equation}
{\psi }_{\kappa}^{3}(x)=\begin{cases}
\mathcal{A} e^{i \kappa x}+\mathcal{B} e^{-i \kappa x},& x<0\\ 

\mathcal{C} e^{i k_5 x}+\mathcal{D} e^{-i k_5 x},& 0<x<1\\

\mathcal{E} e^{i k_6 x}+\mathcal{F} e^{-i k_6 x},& 1<x<2\\

\mathcal{G} e^{i \kappa x},& x>2
\end{cases}.
 \label{eq:eigenstates23}
\end{equation}
For free states, we first obtain an equation from normalization Eq. \ref{eq:orthonormalization}.
\begin{eqnarray*}
&&\int_{-\infty }^{+\infty } \left| \psi _{\kappa }^3(x)\right| {}^2 \, dx \sim \int_{-\infty }^0 \left(| \mathcal{A}| ^2+| \mathcal{B}| ^2\right) \,dx+\int_2^{+\infty } | \mathcal{G}| ^2 \, dx \nonumber\\
&&\sim \int_0^{+\infty } \left(| \mathcal{A}| ^2+| \mathcal{B}| ^2+| \mathcal{G}| ^2\right) \, dx\sim \int_0^{+\infty } \frac{1}{\pi } \, dx,
\end{eqnarray*}
thus,
\begin{equation}
    | \mathcal{A}| ^2+| \mathcal{B}| ^2+| \mathcal{G}| ^2=\frac{1}{\pi }
\end{equation}
Meanwhile, from the probability conservation equation, we obtain:
\begin{equation}
   | \mathcal{G}| ^2=| \mathcal{A}| ^2-| \mathcal{B}| ^2
\end{equation}
From the two equations above, find $\mathcal{A}=\frac{1}{\sqrt{2\pi }}$ (positive real number without loss of generality).We further incorporate the boundary conditions to solve for all coefficients.

\paragraph{Free states 2} For $\varepsilon>0$,the particle is incident from the right, the SE is the same as that in paragraph $c$.
The free states are given by:
\begin{equation}
{\psi }_{\kappa}^{4}(x)=\begin{cases}
\mathcal{G'} e^{-i \kappa x},& x<0\\ 

\mathcal{E'} e^{i k_5 x}+\mathcal{F'} e^{-i k_5 x},& 0<x<1\\

\mathcal{C'} e^{i k_6 x}+\mathcal{D'} e^{-i k_6 x},& 1<x<2\\

\mathcal{A'} e^{i \kappa x}+\mathcal{B'} e^{-i \kappa x},& x>2
\end{cases}.
\label{eq:eigenstates24}
\end{equation}
The normalization equation:
\begin{equation}
    | \mathcal{A'}| ^2+| \mathcal{B'}| ^2+| \mathcal{G'}| ^2=\frac{1}{\pi }.
\end{equation}
The probability conservation equation:
\begin{equation}
   | \mathcal{G'}| ^2=| \mathcal{A'}| ^2-| \mathcal{'B}| ^2.
\end{equation}
From the two equations above, find $\mathcal{A'}=\frac{1}{\sqrt{2\pi }}$. We further incorporate the boundary conditions to solve for all coefficients.

The orthogonality check between these four eigenstates is relatively straightforward, and we omit further details here.

\subsubsection{\label{sec:level122}The completeness of the set of eigenstates}
To simplify the calculations, we choose $V_0=4.27$ and $V_1=1.43$. This ensures that bound states 1  and 2 each have exactly one eigenstate ($\varepsilon_1=-2.80,\varepsilon_2=-0.35 $). 

The measurement probability amplitudes (or projective measurements) of the general initial state (Eq.\ref{eq:initial states}) in the above four cases are:
\begin{equation}
    \varphi _j^1=\int_{\tau }^{\sigma +\tau } \psi _1^1(x){}^* \Psi _j(x) \, dx,
\end{equation}
\begin{equation}
    \varphi _j^2=\int_{\tau }^{\sigma +\tau } \psi _1^2(x){}^* \Psi _j(x) \, dx,
\end{equation}
\begin{equation}
    \varphi _j^3(\kappa )=\int_{\tau }^{\sigma +\tau } \psi _\kappa ^3(x){}^* \Psi _j(x) \, dx,
\end{equation}
\begin{equation}
    \varphi _j^4(\kappa )=\int_{\tau }^{\sigma +\tau } \psi _\kappa ^4(x){}^* \Psi _j(x) \, dx.
\end{equation}
The expansion function of the general initial state in terms of the set of eigenstates is:
\begin{eqnarray}
 &&{f}_{j}(x)={\varphi }_{j}^{1} {\psi }_{1}^{1}(x)+{\varphi }_{j}^{2} {\psi }_{1}^{2}(x)+\int_{0}^{\infty }{\varphi }_{j}^{3}(\kappa ) {\psi }_{\kappa }^{3}(x)\,d\kappa \nonumber\\
&&+\int_{0}^{\infty }{\varphi }_{j}^{4}(\kappa ) {\psi }_{\kappa }^{4}(x)\,d\kappa. 
\label{expansion function 2}
\end{eqnarray}
This equation is the expansion formula for the mixed spectrum. First, we note that the spectral function ${\kappa}(\varepsilon) =\sqrt{2\varepsilon}$ has the dimension of wave number and conforms to the form of Eq. \ref{eq:SF}. Meanwhile, the transformation between Eqs. 18–21 and Eq. 22 corresponds to the representation transformation between coordinate space and momentum space.

We consider two types of initial states, namely $\tau =0.25$, $\sigma =1.63$ (denoted as $j$) and $\tau =-1.17$, $\sigma =4.23$ (denoted as $i$).For these two types of initial state, the completeness of the set of double-well eigenstates holds well (Fig. 1). Interestingly, the first type of initial states is completely within the well, and  the corresponding probability amplitudes are entirely determined by the intrawell quantum number $k_n$. However, the expansion of the continuous spectrum nevertheless employs the seemingly irrelevant quantum number $\kappa$.

\begin{figure}[t]  % t=顶部，b=底部，p=单独一页（PRL推荐的浮动体选项）
  \centering  % 图片居中
  % 关键：width=\linewidth 自动适配单栏宽度，keepaspectratio保持宽高比（避免拉伸）
  \includegraphics[width=\linewidth, keepaspectratio]{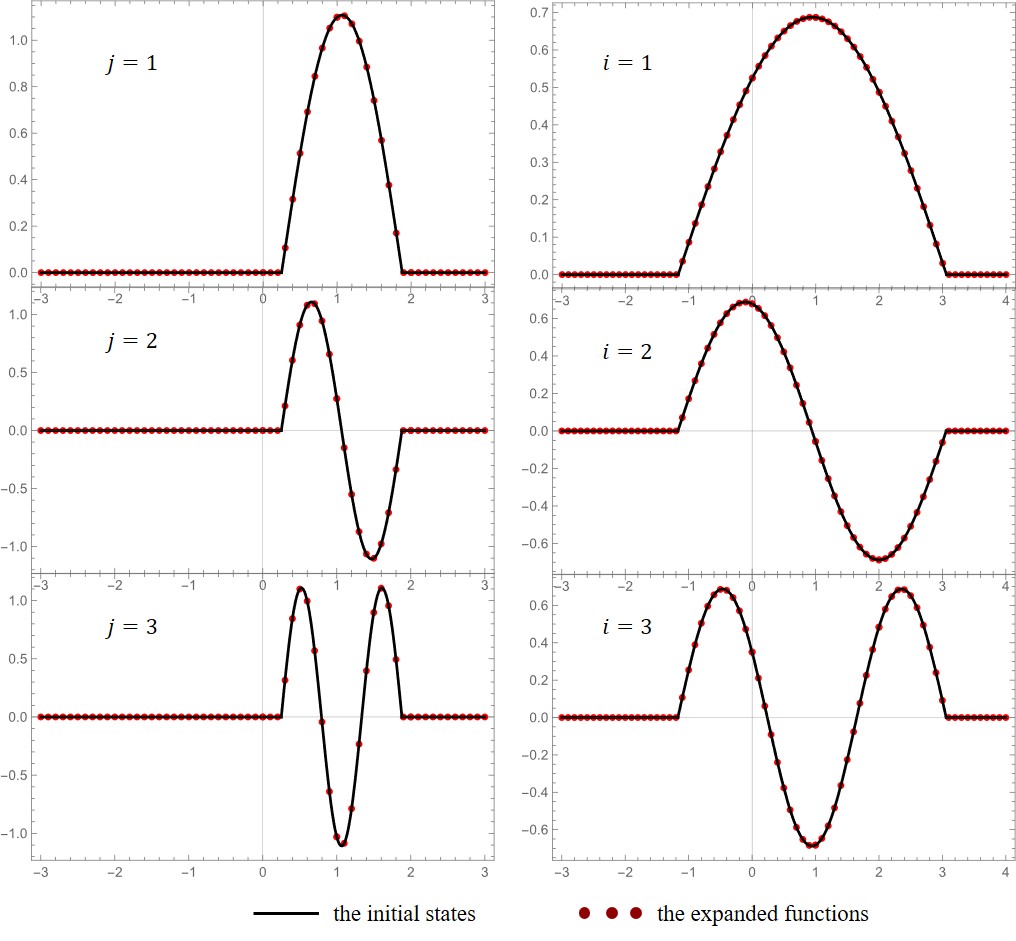}  % 图片文件（矢量图优先）
  \caption{Comparison of initial states $\Psi_n(x)$ and expansion functions $f_n(x)$ of double-well eigenstates. The double-well is located in the region $0 < x < 2$. ($n=j$) Initial states completely within the double-well.  
($n=i$) Initial states spanning both sides of the double-well.}
  \label{fig:double-Well}
\end{figure}

\subsection{\label{sec:level13}Identical finite periodic potential on two-sides}
We do not consider non-periodic oscillatory potentials at infinity, as such potentials are not well-suited to physical requirements, or equivalently, cannot be detected experimentally. Examples of periodic potentials include Bloch electrons \cite{Bloch}, Mathieu equations \cite{Mathieu}, among others. This section discusses two cases: the completeness of eigenstates in a cosine potential and the incompleteness of eigenstates in a discontinuous periodic potential.

\subsubsection{\label{sec:level131}The set of eigenstates in a cosine potential}
We adopt natural units where \( \hbar = m = a =  1 \), the potential:
\begin{equation}
    V(x) = V_0 \cos(2x)
\end{equation}
Here, $V_0$ is a positive real number; for simplicity without loss of generality, we set it to 1. From Eq. \ref{eq:SF} and 23, it is not possible to determine its spectral function.
SE:
\[\psi ''(x)+2 (\varepsilon -\cos (2 x)) \psi (x)=0.\]
Its general solution is given by:
\begin{equation}
    \psi (x)=A \text{MC}[2 \varepsilon,1,x]+B \text{MS}[2 \varepsilon,1,x]
\end{equation}
where $\text{MC}[2 \varepsilon,1,x]$ and $\text{MS}[2 \varepsilon,1,x]$ denote the even and odd Mathieu functions in Wolfram Mathematica, respectively. To uniquely determine the energy eigenstates, we may adopt the Bloch method \cite{Bloch1} (or the Floquet method \cite{Floquet}). By virtue of the potential's periodicity $\pi $, the system possesses discrete translational symmetry, whence we obtain:
\begin{equation}
    \hat{T}\psi \left ( x\right )=\psi \left ( x+\pi \right )={e}^{i\pi \kappa }\psi \left ( x\right )
\end{equation}
Where $e^{i \pi \kappa }$ is the eigenvalue of the translation operator $\hat{T}$, $\kappa(\varepsilon)$ corresponds to the spectral function.
Substituting Eq. 24 into Eq. 25 and enforcing the continuity of $\psi$ and $\psi'$ at $x = \pi$, we obtain:
\begin{equation}
    B=\frac{-\text{MC}[2 \varepsilon,1,\pi ]+e^{i \pi \kappa } \text{MC}[2 \varepsilon ,1,0]}{\text{MS}[2 \varepsilon,1,\pi ]}A
\end{equation}
From the existence of non-trivial Bloch wavefunctions, we derive the conditions that the spectral function $\kappa(\varepsilon)$ must satisfy:
\begin{scriptsize} %  \footnotesize, \scriptsize, \tiny
\begin{equation}
\kappa (\varepsilon )=\pm \frac{1}{\pi } \cos ^{-1}\left(\frac{\text{MC}[2 \varepsilon ,1,\pi ] \text{MS'}[2 \varepsilon ,1,0]+\text{MC}[2 \varepsilon ,1,0] \text{MS'}[2 \varepsilon ,1,\pi ]}{2 \text{MC}[2 \varepsilon ,1,0] \text{MS'}[2 \varepsilon ,1,0]}\right)
\end{equation}
\end{scriptsize}
where $\text{MS'}$ denotes the partial derivative of $\text{MS}$ with respect to $x$. From Eq. 27, we note that $-1 < \kappa \leq 1$ or $-1 \leq \kappa < 1 $. With $\kappa$ restricted to this range, $(\kappa, \varepsilon)$ form a complete set of quantum numbers (with the eigenstates being orthonormal), or equivalently, this corresponds to the use of the reduced zone scheme \cite{ReduceZone}. $\kappa (\varepsilon )$ is the inverse of the band function, whose band structure is illustrated in Fig. 2a-c.

From Eq. 25, we find that ${\left | \psi (x+\pi )\right |}^{2}={\left | \psi (x)\right |}^{2}$; to determine the normalization coefficient $A$, it suffices to apply Eq. \ref{eq:orthonormalization} over one period.
\begin{equation}
    \int_{0}^{\pi }{\left | \psi (x)\right |}^{2}\,dx=\int_{0}^{\pi }{\left | \frac{1}{\sqrt{2\pi }}\right |}^{2}\,dx
\end{equation}
Substituting Eqs. 24 and 26 into the above equation, we find that $A = \frac{1}{\sqrt{2\pi }}$. The shape of the wavefunction is illustrated in Fig. 2d.

\subsubsection{\label{sec:level132}The completeness of the set of eigenstates in a cosine potential}

The measurement probability amplitudes of the general initial state (Eq.\ref{eq:initial states}) in the above two cases are:
\begin{equation}
    \varphi _j^+(\kappa )=\int_{\tau }^{\sigma +\tau } \psi _\kappa ^+(x){}^* \Psi _j(x) \, dx,
\label{The measurement probability amplitudes +}
\end{equation}
\begin{equation}
    \varphi _j^-(\kappa )=\int_{\tau }^{\sigma +\tau } \psi _\kappa ^-(x){}^* \Psi _j(x) \, dx.
\label{The measurement probability amplitudes -}
\end{equation}
Here, $+$ and $-$ correspond to positive and negative values of $\kappa$, respectively.
The expansion function of the general initial state in terms of the set of eigenstates is:
\begin{eqnarray}
 {f}_{j}(x)=\int_{-1}^{0 }{\varphi }_{j}^{-}(\kappa ) {\psi }_{\kappa }^{-}(x)\,d\kappa 
+\int_{0}^{1}{\varphi }_{j}^{+}(\kappa ) {\psi }_{\kappa }^{+}(x)\,d\kappa. 
\label{expansion function 3}
\end{eqnarray}
We take $\tau = 0.15$ and $\sigma = 2.1$, i.e., the initial state is confined to one period. For wider initial states, since a potential with period $a$ is evidently also periodic with $2a, 3a, \ldots$, one can always confine the initial state to a single period. 

From Eq. 27, we find that $\varepsilon(\kappa)$ does not admit an elementary form. Thus, the following change of integration variable is performed in Eq. 31:
\begin{scriptsize} %  \footnotesize, \scriptsize, \tiny
\begin{equation}
{f}_{j}(x)=\int\limits_{0}^{\infty}{\varphi }_{j}^{-}(\kappa ) {\psi }_{\kappa }^{-}(x)\left | \frac{\,d\kappa}{\,d\varepsilon }\right |\,d\varepsilon 
+\int\limits_{0}^{\infty}{\varphi }_{j}^{+}(\kappa ) {\psi }_{\kappa }^{+}(x)\left | \frac{\,d\kappa}{\,d\varepsilon }\right |\,d\varepsilon. 
\label{expansion function 3'}
\end{equation}
\end{scriptsize}
For convenience, it is agreed that the integral sign "$\int\limits_{0}^{\infty }$" denotes the integral excluding discontinuous parts within the integral range, and this convention shall apply to all subsequent contexts.
From Figs. 2a–c, we observe that $\frac{\,d\kappa}{\,d\varepsilon }$ is infinite at the intersection points with the horizontal axis. This singularity is referred to as a Van Hove singularity \cite{VanHove}. As inferred from Eqs. 31 and 32, the integral exists but is difficult to compute accurately. As shown in Figs. 2e–g, for $j=1$, ${f}_{j}(x)$ deviates from $\Psi _j(x)$ by up to 0.012. However, for $j=2$ and $3$, the completeness exhibits excellent behavior. Furthermore, as we will see in the next subsection (the Dirac comb), the spectral function and completeness perform well in periodic potentials. Thus, the deviation for $j=1$ is an algorithmic issue and does not affect our conclusions.

\begin{figure}[b]  % t=顶部，b=底部，p=单独一页（PRL推荐的浮动体选项）
  \centering  % 图片居中
  % 关键：width=\linewidth 自动适配单栏宽度，keepaspectratio保持宽高比（避免拉伸）
  \includegraphics[width=\linewidth, keepaspectratio]{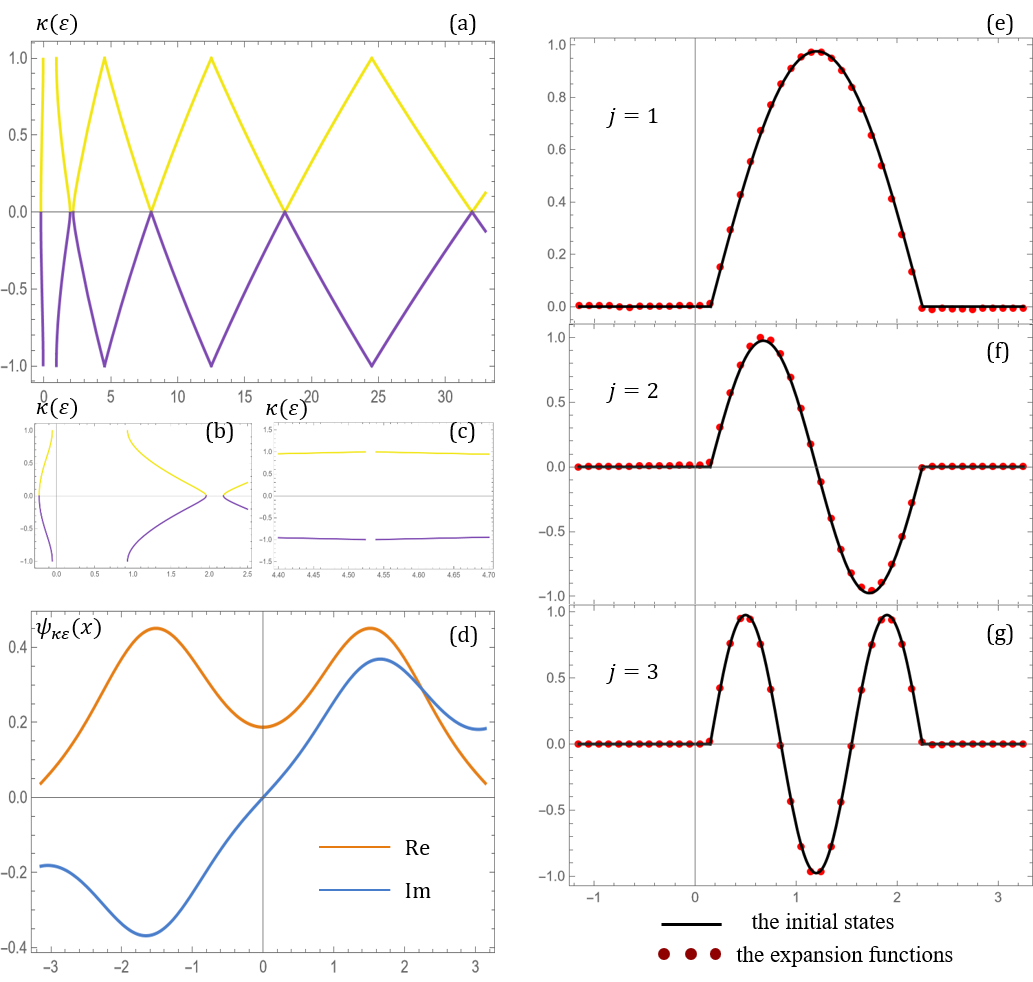}  % 图片文件（矢量图优先）
  \caption{A cosine potential: (a)–(c) Band structure of the spectral function $\kappa (\varepsilon)$; (d) Wavefunction at $\varepsilon= -0.17,\kappa=0.44$; (e)–(j) Comparison between the expansion functions $f_j(x)$ and the initial states $\Psi_j(x)$.}
  \label{fig:Cosine potential}
\end{figure}

\subsubsection{\label{sec:level133}The incompleteness of eigenstates in periodic square potential}
We adopt natural units where \( \hbar = m = a =  1 \), the potential within one period is given by:
\[V\left ( x\right )=\begin{cases}
{V}_{0}, & -b<x<0\\ 
{V}_{1},& 0<x<1
\end{cases}.\]
where $b=0.43,V_0=2.645,V_1=0.27$. For the energy eigenvalue of the system $\varepsilon$,if $V_1 < \varepsilon < V_0$, the spectral function:
For the energy eigenvalue $\varepsilon$ of the system, if $V_1 < \varepsilon < V_0$, the spectral function is given by:
\begin{footnotesize} %  \footnotesize, \scriptsize, \tiny
\begin{equation}
    \kappa (\varepsilon )=\pm \frac{1}{b+1} \cos ^{-1}\left(\cos \beta  \cosh (b \gamma )-\frac{\left(\beta ^2-\gamma ^2\right) \sin \beta  \sinh (b \gamma )}{2 \beta  \gamma }\right)
\end{equation}
\end{footnotesize}
where $\beta =\sqrt{2 (\varepsilon -V_1)},\gamma =\sqrt{2 (V_0-\varepsilon )}.$
if $ \varepsilon > V_0$, the spectral function:
\begin{footnotesize} %  \footnotesize, \scriptsize, \tiny
\begin{equation}
    \kappa (\epsilon )=\pm \frac{1}{b+1} \cos ^{-1}\left(\cos \beta \cos (\alpha  b)-\frac{\left(\alpha ^2+\beta ^2\right) \sin \beta \sin (\alpha  b)}{2 \alpha  \beta }\right)
\end{equation}
\end{footnotesize}
where $\beta =\sqrt{2 (\varepsilon -V_1)},\alpha =\sqrt{2 (\varepsilon -V_0)}.$ From Eqs. 33 and 34, we obtain the complete band-like image of the spectral function, as shown in Fig. 3a.

Solutions to the eigenstates of periodic square potential (Kronig-Penney model) can be found in some quantum mechanics textbooks \cite{periodicsquare}; we do not elaborate on them here. To demonstrate the incompleteness of its eigenstates, it suffices to show that the measurement probability does not normalize to unity—since the normalization of measurement probability is a necessary condition for the completeness of eigenstates \cite{Landau}. The measurement probability amplitudes of the general initial state (Eq.\ref{eq:initial states}) in the above four cases are:
\begin{equation}
    \varphi _j^{i}(\kappa )=\int_{\tau }^{\sigma +\tau } \psi _\kappa ^{i}(x){}^* \Psi _j(x) \, dx.
\end{equation}
Here, $\psi_\kappa^i(x)$ (with $i = 1, 2, 3, 4$) denote the four eigenstates with eigenvalues $(\kappa, \varepsilon)$, classified based on the spectral function. The total measurement probability can be obtained as:
\begin{eqnarray}
    &&{P}_{j}=\int {\left | \varphi _j^{i}(\kappa )\right |}^{2} \,d\kappa=\nonumber\\
&&\int\limits_{V_1}^{V_0} {\left | \varphi _j^{i}(\kappa )\right |}^{2} \left | \frac{d\kappa }{d\varepsilon }\right | \,d\varepsilon+\int\limits_{V_0}^{\infty } {\left | \varphi _j^{i}(\kappa )\right |}^{2} \left | \frac{d\kappa }{d\varepsilon }\right | \,d\varepsilon
\label{eq:the totalmeasurement probability}
\end{eqnarray}
From the range of $\varepsilon$, it follows that the last two integrals in the preceding equation each have two terms.

With $\tau = -0.38$ and $\sigma = 1.25$, the total measurement probability is shown in Table ~\ref{tab:the totalmeasurement probability}. It reveals that the maximum deviation exceeds 0.61, which cannot be attributed to computational errors—indicating the incompleteness of the eigenstate set.

\begin{table}[b]%The best place to locate the table environment is directly after its first reference in text
\caption{\label{tab:the totalmeasurement probability}%
The total measurement probability in periodic square potential.}
\centering
\setlength{\tabcolsep}{3pt} % 默认6pt，减小可增加表格宽度
    \begin{tabular}{ccccccc}
    \hline
    $j$ & 1 & 2 & 3 & 4 & 5 & 6 \\
    $1-P_j$& -0.369 & -0.514 & 0.616 & 0.068 & -0.595 & 0.300  \\
    \hline
    \end{tabular}
\end{table}

\subsubsection{\label{sec:level134}A physically valid initial state}
For the initial state given by Eq. 1, it may be too restrictive physically. Instead, we assume the particle is initially confined in a box, where the potential is continuous except for the left and right walls, i.e.,
\begin{equation}
\mathcal{V}\left ( x\right )=\begin{cases}
+\infty, & x<\tau\\ 
{V}_{0}, & \tau<x<0\\ 
{V}_{1},& 0<x<\tau +\sigma \\
+\infty, & x>\tau +\sigma 
\end{cases}.    
\end{equation}

For the energy eigenstates in $\mathcal{V}(x)$, the spectrum is purely discrete. We select the first excited state therein as the initial state ($\varepsilon=13.728 $) in the aforementioned box.
\begin{equation}
\phi (x)=A\begin{cases}
 0, & x<\tau\\ 
\sin (\xi (x-\tau )), & \tau<x<0\\ 
\frac{\sin (\xi  \tau )}{\sin (\eta(\tau +\sigma) )}\sin (\eta  (\tau +\sigma -x)),& 0<x<\tau +\sigma \\
0, & x>\tau +\sigma    
\end{cases}.
\end{equation}
Here,$\tau=-0.38,\sigma=1.25,A=1.264,\xi=4.708,\eta=5.188.$ The difference between $\phi (x)$ and the first excited state ${\Psi }_{2}(x)$ given by Eq. 1 is shown in Fig. 3b.

Using the eigenstate set from the previous subsubsection to expand the initial state in Eq. 38, we find that the total measurement probability is $P = 0.417$. This indicates the incompleteness of eigenstates in periodic potentials with finite discontinuities, or equivalently, in infinite sets of finite-discontinuity potentials—consistent with the S-L theorem for purely discrete spectra \cite{discrete}. In contrast, as will be shown in the next subsubsection, the eigenstates in the Dirac comb potential exhibit completeness.

\begin{figure}[t]  % t=顶部，b=底部，p=单独一页（PRL推荐的浮动体选项）
  \centering  % 图片居中
  % 关键：width=\linewidth 自动适配单栏宽度，keepaspectratio保持宽高比（避免拉伸）
  \includegraphics[width=\linewidth, keepaspectratio]{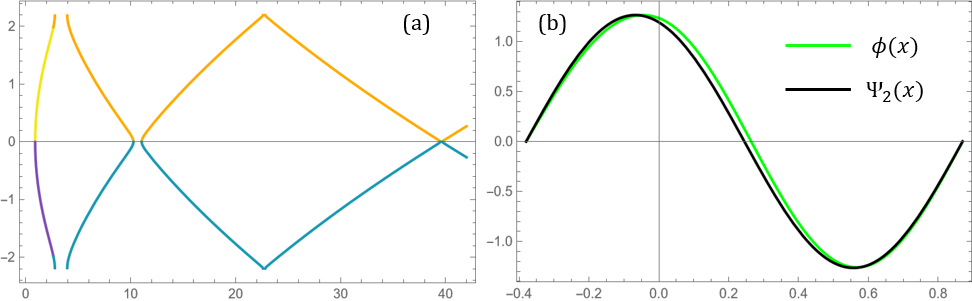}  % 图片文件（矢量图优先）
  \caption{(a)Band structure of the spectral function $\kappa (\varepsilon)$ in periodic square potential; (b) Comparison of two initial states.}
  \label{fig:periodic square potential}
\end{figure}

\subsection{\label{sec:level14}Identical infinite periodic potential on two sides}
\subsubsection{\label{sec:level41}The set of eigenstates in the Dirac comb potential}
The potential within one period is given by:
\begin{equation}
    V(x) = \gamma \delta (x), -a/2<x<a/2.
\end{equation}
We adopt natural units where \( \hbar = m = \gamma =  1 \). 
SE:
\[\psi ''(x)=2 \psi (x) (\delta (x)-\varepsilon ),-a/2<x<a/2.\]
As described in Sec. C1, the solutions to the SE are chosen as the common eigenstates of $(H, \hat{T})$:
\begin{footnotesize} %  \footnotesize, \scriptsize, \tiny\begin{equation}
\begin{equation}
 \psi (x)=\mathcal{A}\begin{cases}
 \mathcal{B} e^{-i k x}+e^{i k x},& -a/2<x<0\\ 
 e^{i a \kappa } \left(\mathcal{B} e^{-i k (x-a)}+e^{i k (x-a)}\right),&0<x<a/2 
\end{cases}.
\end{equation}
\end{footnotesize}
where $k=\sqrt{2\varepsilon },\mathcal{B}=\left(i k e^{i a (\kappa -k)}-i k-1\right)$. An apparent dimensional inconsistency arises in $\mathcal{B}$, which stems from the adoption of natural units and the $\delta$-potential jump condition. It is merely a formal issue with no impact on the results. To determine the normalization coefficient $\mathcal{A}$, it suffices to apply Eq. \ref{eq:orthonormalization} over one period.
\begin{equation}
    \int_{-a/2}^{a/2 }{\left | \psi (x)\right |}^{2}\,dx=\int_{-a/2}^{a/2}{\left | \frac{1}{\sqrt{2\pi }}\right |}^{2}\,dx,
\end{equation}
 we find that:
\begin{equation}
\mathcal{A}=\sqrt{\frac{a/(2 \pi )}{A-B}
},
\end{equation}
\[A=1+2 a \left(k^2+1\right)+\cos (a (k+\kappa))-\cos (2 a k),\]
\[B=\left(2 a k^2+1\right) \cos (a (k-\kappa ))+2 a k \sin (a (k-\kappa ))+{\sin (2 a k)}/{k}.\]
The spectral function $\kappa(\varepsilon)$:
\begin{equation}
\kappa (k)=\pm \frac{1}{a} \cos ^{-1}\left(\sin (a k)/{k}+\cos (a k)\right)
\end{equation}

With $a = 1.3$, the band-like structure images of $\kappa(\varepsilon)$ and $\kappa(k)$ are shown in Figs. 4a and 4b, respectively, and the wavefunction image in Fig. 4c.

\subsubsection{\label{sec:level142}The completeness of the set of eigenstates in the Dirac comb potential}
The measurement probability amplitudes of the general initial state (Eq.\ref{eq:initial states}) are identical to Eqs. 29 and 30.
The expansion function of the general initial state in terms of the set of eigenstates is identical to Eqs. \ref{expansion function 3} or \ref{expansion function 3'}. Here, one can also choose $k$ as the independent variable, with the form given below:
\begin{scriptsize} %  \footnotesize, \scriptsize, \tiny
\begin{equation}
{f}_{j}(x)=\int\limits_{0}^{\infty}{\varphi }_{j}^{-}(\kappa ) {\psi }_{\kappa }^{-}(x)\left | \frac{\,d\kappa}{\,dk}\right |\,dk 
+\int\limits_{0}^{\infty}{\varphi }_{j}^{+}(\kappa ) {\psi }_{\kappa }^{+}(x)\left | \frac{\,d\kappa}{\,dk}\right |\,dk. 
\label{expansion function 4}
\end{equation}
\end{scriptsize}
This form can accelerate the speed of numerical computations.

We take $\tau = 0.45$ and $\sigma = 1 $, i.e., the initial state is confined to one period. As shown in Figs. 4d–f, the completeness of the eigenstate set in the Dirac comb potential is well-behaved.

Sections IIIC and IIID discuss two types of periodic potentials, respectively. In fact, the SE for any periodic potential is a Hill equation:
\[w''(z)+\left ( \lambda +Q(z)\right )w(z)=0,Q(z+a)=Q(z).\]
Here,$a$ is the period. Many properties of solutions to the Mathieu equation are shared by solutions to the Hill equation\cite{Hill}.

\begin{figure}[t]  % t=顶部，b=底部，p=单独一页（PRL推荐的浮动体选项）
  \centering  % 图片居中
  % 关键：width=\linewidth 自动适配单栏宽度，keepaspectratio保持宽高比（避免拉伸）
  \includegraphics[width=\linewidth, keepaspectratio]{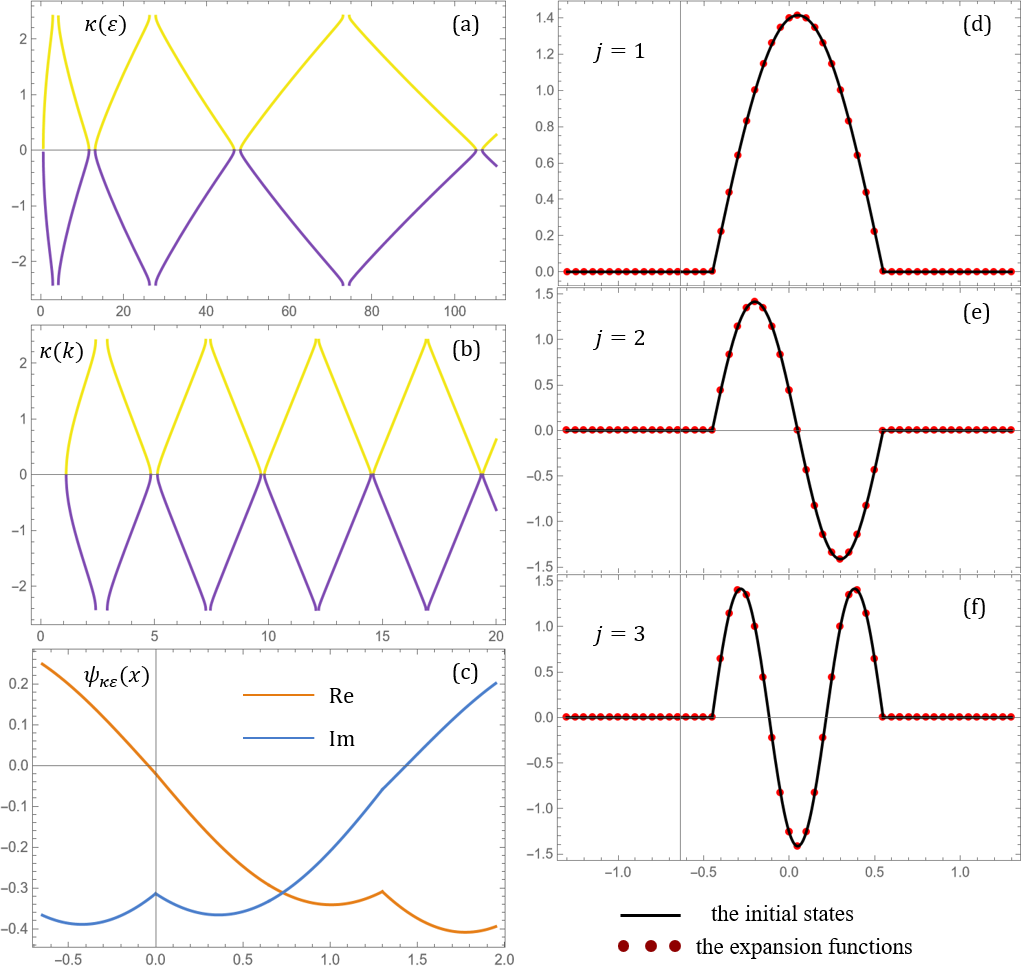}  % 图片文件（矢量图优先）
  \caption{Dirac comb potential: (a)–(b) Band structure of the spectral function $\kappa (\varepsilon)$; (c) Wavefunction at $k=1.5,\kappa=-1.01$; (d)–(f) Comparison between the expansion functions $f_j(x)$ and the initial states $\Psi_j(x)$.}
  \label{fig:Dirac comb potential}
\end{figure}

\section{\label{sec:level2}Part Two: Asymmetric potential}
In asymmetric potentials, where translational symmetry is broken, finding the spectral function $\kappa(\varepsilon)$ — the original integration variable (see Eqs. \ref{expansion function 2}, \ref{expansion function 3}, \ref{expansion function 3'}, etc.) — becomes intricate, except in a few cases (see Subsects. B and D below).
\subsection{\label{sec:level21}Unequal constant potential on two sides}
We will present a rigorous theoretical proof for the completeness of the eigenstate set in this potential. Such a potential is apparently simple yet inherently complex, with its general form given by:
\begin{equation}
    \mathcal{V}(x)=\begin{cases}
 -V_0/2,&x<0 \\ 
V_0/2, & x>0
\end{cases}.
\end{equation}
Here $V_0>0$. Taking $-V_0/2$ as the potential energy reference plane, the potential function becomes the following step potential:
\begin{equation}
V\left ( x\right )=\begin{cases}
0 ,& x<0 \\ 
V_0 ,& x>0
\end{cases}.
\end{equation}
We now discuss the step potential as an example. 
\subsubsection{\label{sec:level211}The set of eigenstates in unequal constant potential}
\paragraph{Free states 1} For energy eigenvalue $0<\varepsilon<V_0$, the SE:
\[x<0 ,{\psi }''(x)+k^{2}\psi(x) =0,k=\sqrt{2\varepsilon };\]
\[x>0 ,{\psi }''(x)-\alpha^{2}\psi(x) =0,\alpha =\sqrt{2 (V_0-\varepsilon )}.\]
From the finiteness of $\psi(x)$ as $x \to \infty$, we obtain the free states as:
\begin{equation}
\psi^0 (x)=\begin{cases}
 \mathcal{A} e^{i k x}+\mathcal{B}e^{-i k x}, &x<0 \\ 
 \mathcal{C} e^{-\alpha x}, &x>0 
\end{cases}.
\end{equation}
For free states, we obtain an equation from normalization Eq. \ref{eq:orthonormalization}.
\begin{footnotesize} %  \footnotesize, \scriptsize, \tiny
\begin{eqnarray}
&&\int_{-\infty }^{+\infty } \left| \psi^0(x)\right| {}^2 \, dx \sim \int_{-\infty }^0 \left(| \mathcal{A}| ^2+| \mathcal{B}| ^2\right) \,dx+\int_{0}^{+\infty } | \mathcal{C}| ^2 \, dx \nonumber\\
&&\sim \int_{-\infty}^{0} \left(| \mathcal{A}| ^2+| \mathcal{B}| ^2\right) \, dx\sim \int_{-\infty}^{0} \frac{1}{\pi } \, dx,
\end{eqnarray}
\end{footnotesize}
thus,
\begin{equation}
    | \mathcal{A}| ^2+| \mathcal{B}| ^2=\frac{1}{\pi }
\end{equation}
Meanwhile, from continuity of $\psi(x)$ and $\psi'(x)$ at $x = 0$, we obtain:
\begin{equation}
  \mathcal{A}=\frac{1}{2} \left(1+\frac{i \alpha }{k}\right) \mathcal{C},
\end{equation}
\begin{equation}
   \mathcal{B}=\frac{1}{2} \left(1-\frac{i \alpha }{k}\right) \mathcal{C}.
\end{equation}
From the three equations above, we find that:
\begin{eqnarray*}
&&\mathcal{C}=\sqrt{{2 k^2}/(\pi \left(\alpha ^2+k^2\right))}\nonumber\\
&&\mathcal{A}=\mathcal{C},\mathcal{B}=\sqrt{{2 \alpha ^2}/(\pi \left(\alpha ^2+k^2\right))}
\end{eqnarray*}

\paragraph{Free states 2} For energy eigenvalue $\varepsilon>V_0$, the SE:
\[x<0 ,{\psi }''(x)+k^{2}\psi(x) =0,k=\sqrt{2\varepsilon };\]
\[x>0 ,{\psi }''(x)+\beta^{2}\psi(x) =0,\beta =\sqrt{2 (\varepsilon-V_0 )}.\]

Up to this point, the traditional approach—as outlined in Secs. IIIBc and IIIBd—involves partitioning the eigenstates into left-incident $\psi^l(x)$ and right-incident $\psi^r(x)$. However, due to the asymmetry between the left and right potentials in the present case, $\psi^l(x)$ and $\psi^r(x)$ are non-orthogonal. Physically, non-orthogonality implies that these two states are indistinguishable, thus the eigenstates obtained via this approach fail to meet the measurement requirements. The detailed derivation is left to the interested reader for verification and is not presented here.

From the finiteness of $\psi(x)$ as $x \to \infty$, we obtain the free states as:
\begin{equation}
\psi (x)=\begin{cases}
\mathcal{D} \cos (k x)+\mathcal{E} \sin (k x), &x<0 \\ 
\mathcal{F}\cos (\beta  x)+\mathcal{G} \sin (\beta  x), &x>0 
\end{cases}.
\end{equation}
For free states, we obtain an equation from normalization Eq. \ref{eq:orthonormalization}.
\begin{scriptsize} %  \footnotesize, \scriptsize, \tiny
\begin{eqnarray}
&&\int_{-\infty }^{+\infty } \left| \psi(x)\right| {}^2 \, dx \sim \frac{1}{2} \int_{-\infty }^0 \left(| \mathcal{D}| ^2+| \mathcal{E}| ^2\right) \,dx+\frac{1}{2} \int_{0}^{+\infty } (|\mathcal{F}| ^2+|\mathcal{G}| ^2 )\, dx \nonumber\\
&&\sim \frac{1}{2}\int_{-\infty}^{0} \left(| \mathcal{D}| ^2+| \mathcal{E}| ^2+|\mathcal{F}| ^2+|\mathcal{G}| ^2\right) \, dx\sim \int_{-\infty}^{0} \frac{1}{\pi } \, dx,
\end{eqnarray}
\end{scriptsize}
thus,
\begin{equation}
    | \mathcal{D}| ^2+| \mathcal{E}| ^2+|\mathcal{F}| ^2+|\mathcal{G}| ^2=\frac{2}{\pi }
\end{equation}
Meanwhile, from continuity of $\psi(x)$ and $\psi'(x)$ at $x = 0$, we obtain:
\begin{equation}
  \mathcal{D}=\mathcal{F},
\end{equation}
\begin{equation}
   \mathcal{E}=\frac{\beta}{k}\mathcal{G}.
\end{equation}

At this stage, we find four undetermined coefficients but only three equations. An additional orthonormality equation is required to uniquely determine the eigenstates. Assuming two eigenstates $\psi^1(x)$ and $\psi^2(x)$ with the same energy, characterized by coefficients $(\mathcal{D}, \mathcal{E}, \mathcal{F}, \mathcal{G})$ and $(\mathcal{D}', \mathcal{E}', \mathcal{F}', \mathcal{G}')$ respectively, substituting them into the orthonormality equation (Eq. 2) yields:
\begin{scriptsize} %  \footnotesize, \scriptsize, \tiny
\begin{eqnarray}
&&\int_{-\infty }^{\infty } \left(\psi ^{1}(x)\right){}^* \psi ^{2}(x) \, dx\sim \frac{1}{2} \int_{-\infty }^0 \left(\mathcal{D^*}\mathcal{D'}+\mathcal{E^*}\mathcal{E'}\right) \,dx+\frac{1}{2} \int_{0}^{+\infty}\nonumber\\
&&\left(\mathcal{F^*}\mathcal{F'}+\mathcal{G^*}\mathcal{G'}\right)\, dx\sim\frac{1}{2}\int_{0}^{+\infty}\left(\mathcal{D^*}\mathcal{D'}+\mathcal{E^*}\mathcal{E'}+\mathcal{F^*}\mathcal{F'}+\mathcal{G^*}\mathcal{G'}\right)\, dx\nonumber\\
&&=0,
\end{eqnarray}
\end{scriptsize}
thus,
\begin{equation}
\mathcal{D^*}\mathcal{D'}+\mathcal{E^*}\mathcal{E'}+\mathcal{F^*}\mathcal{F'}+\mathcal{G^*}\mathcal{G'}=0
\end{equation}
Substituting Eqs. 55 and 56—corresponding to the coefficients of the two energy eigenstates—into their respective Eqs. 54 and 58 yields the following three equations:
\begin{equation}
    2 | \mathcal{F}| ^2+\frac{\left(\beta ^2+k^2\right) }{k^2}|\mathcal{G}| ^2=\frac{2}{\pi }
\end{equation}
\begin{equation}
    2 | \mathcal{F'}| ^2+\frac{\left(\beta ^2+k^2\right) }{k^2}|\mathcal{G'}| ^2=\frac{2}{\pi }
\end{equation}
\begin{equation}
    2\mathcal{F^*} \mathcal{F'}+\frac{\beta ^2+k^2}{k^2}\mathcal{G^*}\mathcal{G'}=0
\end{equation}
Without loss of generality, we take $\mathcal{F}$ and $\mathcal{F}'$ as non-negative real numbers, and set $\mathcal{G} = G e^{i\theta}$ and $\mathcal{G}' = G' e^{i\Omega}$.  Here, $G, G', \theta,$ and $\Omega$ are all non-negative real numbers. Substituting these into the three equations above yields:
\begin{equation}
    2 \mathcal{F}^2+\frac{ \left(\beta ^2+k^2\right)}{k^2} G^2=\frac{2}{\pi }
\end{equation}
\begin{equation}
    2 \mathcal{F'}^2+\frac{ \left(\beta ^2+k^2\right)}{k^2} G'^2=\frac{2}{\pi }
\end{equation}
\begin{equation}
    2 \mathcal{F} \mathcal{F}'+\frac{\beta ^2+k^2}{k^2} G G' e^{i (\Omega -\theta )} =0
\label{ orthonormality equatio}
\end{equation}
From Eq. 64, we obtain $e^{i(\Omega - \theta)} = -1$, i.e., $e^{i\Omega} = -e^{i\theta}$. Combining Eqs. 62–64 and Eqs. 55–56 , we yield the two orthonormal energy eigenstates as:
\begin{equation}
{\psi }^{1}(x)=\begin{cases}
\mathcal{F} \cos (k x)+\sqrt{\frac{2 \beta ^2 \left(1-\pi  \mathcal{F}^2\right)}{\pi  \left(\beta ^2+k^2\right)}} e^{i \theta } \sin (k x), & x<0 \\ 
\mathcal{F} \cos (\beta  x)+\sqrt{\frac{2 k^2 \left(1-\pi  \mathcal{F}^2\right)}{\pi  \left(\beta ^2+k^2\right)}} e^{i \theta } \sin (\beta  x), &x>0 
\end{cases};
\label{eigenstates 211}
\end{equation}

\begin{equation}
{\psi }^{2}(x)=\begin{cases}
\sqrt{\frac{1-\pi  \mathcal{F}^2}{\pi }} \cos (k x)-\sqrt{\frac{2 \beta ^2 \mathcal{F}^2}{\beta ^2+k^2}} e^{i \theta } \sin (k x), & x<0 \\ 
\sqrt{\frac{1-\pi  \mathcal{F}^2}{\pi }} \cos (\beta  x)- \sqrt{\frac{2 k^2 \mathcal{F}^2}{\beta ^2+k^2}} e^{i \theta } \sin (\beta  x), &x>0 
\end{cases}.
\label{eigenstates 212}
\end{equation}
where $\mathcal{F}$ and $\theta$ are arbitrary non-negative real numbers. Below, we reproduce the free states 1 from the previous subsection for summary:
\begin{equation}
{\psi }^{0}(x)=\begin{cases}
\sqrt{\frac{2 k^2}{\pi  \left(\alpha ^2+k^2\right)}} \cos (k x)-\sqrt{\frac{2 \alpha ^2}{\pi  \left(\alpha ^2+k^2\right)}} \sin (k x), & x<0 \\ 
\sqrt{\frac{2 k^2}{\pi  \left(\alpha ^2+k^2\right)}} e^{-\alpha  x}, &x>0 
\end{cases}.
\label{eigenstates 210}
\end{equation}

\subsubsection{\label{sec:level212}The spectral function $\kappa (\varepsilon)$ in unequal constant potential}
Since no operator commuting with the Hamiltonian $H$ exists, $\psi^1(x)$ and $\psi^2(x)$ remain incompletely determined. This subsection aims to determine $\kappa (\varepsilon)$ and fix the values of $\mathcal{F}$ and $\theta$ via the normalization of the total measurement probability.

Taking the ground state ($j=1$) from Eq. 1 as the initial state, we obtain the measurement probability amplitude:
\begin{equation}
    \varphi ^{i}=\int_{\tau }^{\sigma +\tau } \psi ^{i}(x){}^* \Psi (x) \, dx.
\end{equation}
Here,$i=0,1,2$. The total measurement probability can be obtained as:
\begin{equation}
{P}=\int {\left | \varphi ^{i}\right |}^{2} \,d\mathcal{\kappa }_i
\end{equation}
Different energy eigenvalues $\varepsilon$ correspond to distinct spectral functions $\mathcal{\kappa }_i(\varepsilon)$. This is expected, as in Sec. IIIB, which includes both discrete and continuous spectra. For $0 < \varepsilon < V_0$, the right side exhibits bound behavior, and the spectral functions depend only on the left side. From Eq. 5, we have $\mathcal{\kappa }_0 = \sqrt{2(\varepsilon - 0)} = k$. For $\varepsilon > V_0$, the left-side spectral function is $k$ and the right-side one is $\beta$, with $k \neq \beta$ unless $V_0 = 0$. Thus, for $\varepsilon > V_0$, the spectral functions may be functions of both. Since $\beta = \sqrt{k^2 + V_0}$, we have $\,d\mathcal{\kappa }_i = \frac{\,d\mathcal{\kappa }_i}{\,dk} \,dk$. Under the normalization of the total measurement probability, Eq. 69 expands to:
\begin{eqnarray}
&&1=\int_{0}^{\sqrt{2V_0}} {\left | \varphi ^{0}\right |}^{2} \,dk+\int_{\sqrt{2V_0}}^{\infty } {\left | \varphi ^{1}\right |}^{2} \left |\frac{\,d\mathcal{\kappa }_1}{\,dk}\right | \,dk \nonumber\\
&&+ \int_{\sqrt{2V_0}}^{\infty } {\left | \varphi ^{2}\right |}^{2} \left |\frac{\,d\mathcal{\kappa }_2}{\,dk}\right | \,dk
\end{eqnarray}
Suppose $\tau < 0$, $\sigma > 0$, and $\tau + \sigma < 0$ — i.e., the initial state is entirely confined to the region $x < 0$. Next, fixing the initial state width $\sigma$ and taking the limit $\tau \to -\infty$ in Eq. 70, we apply the Riemann-Lebesgue lemma \cite{Riemann} to obtain the following equation:
\begin{scriptsize} %  \footnotesize, \scriptsize, \tiny
\begin{eqnarray}
&&1=\int_0^{\sqrt{2 V_0}} \frac{8 \pi  \sigma  \cos ^2\left(\frac{k \sigma }{2}\right)}{\left(\pi ^2-k^2 \sigma ^2\right)^2} \, dk+\nonumber\\
&&\int_{\sqrt{2 V_0}}^{\infty } \frac{4 \pi  \sigma  \cos ^2\left(\frac{k \sigma }{2}\right) \left(2 \beta ^2-\pi  \beta ^2 \mathcal{F}^2+\pi  k^2 \mathcal{F}^2\right)}{\left(\beta ^2+k^2\right) \left(\pi ^2-k^2 \sigma ^2\right)^2} \left | \frac{\,d\mathcal{\kappa }_1}{\,dk}\right | \, dk+\nonumber\\
&&\int_{\sqrt{2 V_0}}^{\infty } \frac{4 \pi  \sigma  \cos ^2\left(\frac{k \sigma }{2}\right) \left(\beta ^2+k^2+\pi  \beta ^2 \mathcal{F}^2-\pi  k^2 \mathcal{F}^2\right)}{\left(\beta ^2+k^2\right) \left(\pi ^2-k^2 \sigma ^2\right)^2} \left | \frac{\,d\mathcal{\kappa }_2}{\,dk}\right | \, dk.
\end{eqnarray}
\end{scriptsize}
Next, suppose $\tau > 0$, $\sigma > 0$, and $\tau + \sigma > 0$ — i.e., the initial state is entirely confined to the region $x > 0$. Fixing the initial state width $\sigma$ and taking the limit $\tau \to +\infty$ in Eq. 70, we reapply the Riemann-Lebesgue lemma to obtain the following equation:
\begin{scriptsize} %  \footnotesize, \scriptsize, \tiny
\begin{eqnarray}
&&1=\int_{\sqrt{2 V_0}}^{\infty } \frac{4 \pi  \sigma  \cos ^2\left(\frac{\beta  \sigma }{2}\right) \left(2 k^2+\pi  \beta ^2 \mathcal{F}^2-\pi  k^2 \mathcal{F}^2\right)}{\left(\pi ^2-\beta ^2 \sigma ^2\right)^2 \left(\beta ^2+k^2\right)} \left | \frac{\,d\mathcal{\kappa }_1}{\,dk}\right | \, dk+\nonumber\\
&&\int_{\sqrt{2 V_0}}^{\infty } \frac{4 \pi  \sigma  \cos ^2\left(\frac{\beta  \sigma }{2}\right) \left(\beta ^2+k^2-\pi  \beta ^2 \mathcal{F}^2+\pi  k^2 \mathcal{F}^2\right)}{\left(\pi ^2-\beta ^2 \sigma ^2\right)^2 \left(\beta ^2+k^2\right)} \left | \frac{\,d\mathcal{\kappa }_2}{\,dk}\right | \, dk.
\end{eqnarray}
\end{scriptsize}
Changing the integration variable in Eq. 72 via $\,dk = \frac{\,dk}{\,d\beta} \,d\beta = \frac{\beta}{k} \,d\beta$, we obtain:
\begin{scriptsize} %  \footnotesize, \scriptsize, \tiny
\begin{eqnarray}
&&1=\int_{0}^{\infty } \frac{4 \pi  \sigma  \cos ^2\left(\frac{\beta  \sigma }{2}\right) \left(2 k^2+\pi  \beta ^2 \mathcal{F}^2-\pi  k^2 \mathcal{F}^2\right)}{\left(\pi ^2-\beta ^2 \sigma ^2\right)^2 \left(\beta ^2+k^2\right)} \left | \frac{\,d\mathcal{\kappa }_1}{\,dk}\right | \frac{\beta}{k} \, d\beta+\nonumber\\
&&\int_{0}^{\infty } \frac{4 \pi  \sigma  \cos ^2\left(\frac{\beta  \sigma }{2}\right) \left(\beta ^2+k^2-\pi  \beta ^2 \mathcal{F}^2+\pi  k^2 \mathcal{F}^2\right)}{\left(\pi ^2-\beta ^2 \sigma ^2\right)^2 \left(\beta ^2+k^2\right)} \left | \frac{\,d\mathcal{\kappa }_2}{\,dk}\right | \frac{\beta}{k} \, d\beta.
\end{eqnarray}
\end{scriptsize}

We introduce an integral equation, which can be directly verified or proven using the method from Subsubsection 3. The integral equation reads:
\begin{equation}
  \int_0^{\infty } \frac{8 \pi  \sigma  \cos ^2\left(\frac{k \sigma }{2}\right)}{\left(\pi ^2-k^2 \sigma ^2\right)^2} \, dk=1  
\end{equation}
Comparing Eqs. 71 and 73 with Eq. 74 respectively, we find that the sufficient conditions for their validity are:
\begin{eqnarray}
&&\frac{2 \beta ^2-\pi  \beta ^2 \mathcal{F}^2+\pi  k^2 \mathcal{F}^2}{\beta ^2+k^2}\left | \frac{\,d\mathcal{\kappa }_1}{\,dk}\right |+\nonumber\\
&&\frac{\beta ^2+k^2+\pi  \beta ^2 \mathcal{F}^2-\pi  k^2 \mathcal{F}^2}{\beta ^2+k^2}\left | \frac{\,d\mathcal{\kappa }_2}{\,dk}\right |=2,
\end{eqnarray}

\begin{eqnarray}
&&\frac{\beta  \left(2 k^2+\pi  \beta ^2 \mathcal{F}^2-\pi  k^2 \mathcal{F}^2\right)}{k \left(\beta ^2+k^2\right)}\left | \frac{\,d\mathcal{\kappa }_1}{\,dk}\right |+\nonumber\\
&&\frac{\beta  \left(\beta ^2+k^2-\pi  \beta ^2 \mathcal{F}^2+\pi  k^2 \mathcal{F}^2\right)}{k \left(\beta ^2+k^2\right)}\left | \frac{\,d\mathcal{\kappa }_2}{\,dk}\right |=2.
\end{eqnarray}
Combining Eqs. 75 and 76, we yield:
\begin{equation}
    \left | \frac{\,d\mathcal{\kappa }_1}{\,dk}\right |=\frac{\beta ^2+k^2-\pi  \mathcal{F}^2 (\beta +k)^2}{ \left(1-2 \pi  \mathcal{F}^2\right) \beta (\beta +k)}>0,
\end{equation}
\begin{equation}
    \left | \frac{\,d\mathcal{\kappa }_2}{\,dk}\right |=\frac{2 \beta  k-\pi  \mathcal{F}^2 (\beta +k)^2}{\left(1-2 \pi  \mathcal{F}^2\right) \beta (\beta +k)}>0.
\end{equation}
A sufficient condition for the validity of the two preceding equations is either one of the following two:
\begin{equation}
\pi  \mathcal{F}^2>\frac{\beta ^2+k^2}{(\beta +k)^2}\to \pi  \mathcal{F}^2=1,
\end{equation}
\begin{equation}
\pi \mathcal{F}^2<\frac{2 \beta  k}{(\beta +k)^2}\to \pi \mathcal{F}^2=0.
\end{equation}
The two conditions above effectively yield the same set of equations. Using the first condition, we obtain:
\begin{equation}
    \left | \frac{\,d\mathcal{\kappa }_1}{\,dk}\right |=\frac{2 \beta  k}{\beta  (\beta +k)},
\end{equation}
\begin{equation}
    \left | \frac{\,d\mathcal{\kappa }_2}{\,dk}\right |=\frac{\beta ^2+k^2}{\beta  (\beta +k)}.
\end{equation}
Integrating the two preceding equations and incorporating $\kappa_0$, we obtain the continuous spectral function:
\[\kappa_0=k,\]
\[\kappa_1=\frac{(\beta +k)^2+\left(\beta ^2+k^2\right)}{3 (\beta +k)}+\frac{\sqrt{2 V_0}}{3},\]
\begin{equation}
\kappa_2=-\frac{(\beta +k)^2+2 \beta  k}{3 (\beta +k)}+\frac{\sqrt{2 V_0}}{3}. 
\end{equation}
The structure of the spectral functions is illustrated in Figs. 5a or 5b. Although the spectral function here does not exhibit regional discontinuities as in a periodic potential, eigenstates do not exist at the junction points ($\varepsilon = 0, V_0$). This is because at these junction points, either particles are absolutely forbidden or the states diverge at infinity. The absence of states at junction points in the spectral function is a universal issue in quantum mechanics. Substituting Eq. 79 into Eqs. 65 and 66, we obtain the unique orthonormal energy eigenstates (up to a phase factor) for $\varepsilon > V_0$:
\begin{equation}
{\psi }^{1}(x)=\begin{cases}
\frac{1}{\sqrt{\pi }} \cos (k x), & x<0 \\ 
\frac{1}{\sqrt{\pi }} \cos (\beta  x), &x>0 
\end{cases};
\label{eigenstates 211'}
\end{equation}

\begin{equation}
{\psi }^{2}(x)=\begin{cases}
\sqrt{\frac{2 \beta ^2}{\pi  \left(\beta ^2+k^2\right)}} \sin (k x), & x<0 \\ 
\sqrt{\frac{2 k^2}{\pi  \left(\beta ^2+k^2\right)}} \sin (\beta  x), &x>0 
\end{cases}.
\label{eigenstates 212'}
\end{equation}

\subsubsection{\label{sec:level213}The completenes of the set of eigenstates in unequal constant potential}
This part of the proof requires certain knowledge of complex analysis. First, we prove the completeness of the initial state confined to the region $x < 0$ or $x > 0$. Then, we extend the proof to the case where the initial state involves $x = 0$.
\paragraph{The initial state confined to the region $x < 0$ or $x > 0$}
Here, we take the initial state (Eq. 1) confined to the region $x > 0$ (setting $\tau = 0$) with even $j$ as an example, with the proof for the other three cases being entirely analogous. By combining this initial state with the three types of eigenstates (Eqs. 67, 84, and 85), we obtain the expansion function as($0<x<\sigma $):
\begin{scriptsize} %  \footnotesize, \scriptsize, \tiny
\begin{eqnarray}
&&{f}_{j}(x)=\int {\varphi }_{j}^{i} {\psi }^{i}(x) \left | \frac{\,d{\kappa }_{i}}{\,dk}\right | \,dk\nonumber\\
&&=\int_0^{\sqrt{2 V_0}} \frac{2 \sqrt{2 \sigma } j e^{-\alpha  (x+\sigma)} \left(e^{\alpha  \sigma }-1\right) k^2}{ \left(\alpha ^2 \sigma ^2+\pi ^2 j^2\right) \left(\alpha ^2+k^2\right)} \, dk+\nonumber\\
&&\int_{\sqrt{2 V_0}}^{+\infty } \frac{2 \sqrt{2 \sigma } j k (\beta  (1-\cos (\beta  \sigma )) \cos (\beta  x)-k \sin (\beta  \sigma ) \sin (\beta  x))}{ \left(\pi ^2 j^2-\beta ^2 \sigma ^2\right)\beta  (\beta +k)} \, dk.\nonumber\\
\end{eqnarray}
\end{scriptsize}
Substituting $\alpha = \sqrt{2V_0 - k^2}$ and $\beta = \sqrt{k^2 - 2V_0}$ into the preceding equation and rearranging, we obtain:
\begin{eqnarray}
&&{f}_{j}(x)=\int_{0}^{\sqrt{2V_0}}\frac{j \sqrt{2 \sigma }}{\pi ^2 j^2+2 \sigma ^2 V_0-\sigma ^2 k^2 }{J}_{1}(k)\,dk+\nonumber\\
&&\int_{\sqrt{2V_0}}^{\infty }\frac{j \sqrt{2 \sigma }}{\pi ^2 j^2+2 \sigma ^2 V_0-\sigma ^2 k^2 }{J}_{2}(k)\,dk,
\end{eqnarray}

\begin{footnotesize} %  \footnotesize, \scriptsize, \tiny
\[J_1(k)=-Re\left(\frac{i k \sqrt{2 V_0-k^2} e^{- \sqrt{2 V_0-k^2} (\sigma -x)}}{k^2-2 V_0}\right)+\]
\[Re\left(\frac{i k \sqrt{2 V_0-k^2} \left(i k \sqrt{2 V_0-k^2}-k^2+2 V_0\right) e^{- x \sqrt{2 V_0-k^2}}}{V_0 \left(k^2-2 V_0\right)}\right)-\]
\[Re\left(\frac{i k \sqrt{2 V_0-k^2} \left(i k \sqrt{2 V_0-k^2}-k^2+V_0\right) e^{- \sqrt{2 V_0-k^2} (\sigma +x)}}{V_0 \left(k^2-2 V_0\right)}\right),\]

\[J_2(k)=-Re\left(\frac{k \sqrt{k^2-2 V_0} e^{i \sqrt{k^2-2 V_0} (\sigma -x)}}{k^2-2 V_0}\right)+\]
\[Re\left(\frac{k \sqrt{k^2-2 V_0} \left(k \sqrt{k^2-2 V_0}-k^2+2 V_0\right) e^{i x \sqrt{k^2-2 V_0}}}{V_0 \left(k^2-2 V_0\right)}\right)-\]
\[Re\left(\frac{k \sqrt{k^2-2 V_0} \left(k \sqrt{k^2-2 V_0}-k^2+V_0\right) e^{i \sqrt{k^2-2 V_0} (\sigma +x)}}{V_0 \left(k^2-2 V_0\right)}\right).\]
\end{footnotesize}
The integration path of Eq. 87 is a closed contour $\Gamma$ in the complex plane, illustrated in Fig. 5c. We choose the single-valued holomorphic branch such that both $\sqrt{2V_0 - k^2}$ and $\sqrt{k^2 - 2V_0}$ in Eq. 87 take non-negative real values. The argument range $0 \leq \arg(z) \leq 2\pi$, the calculation method is shown in Fig. 5d. With this choice, we obtain:
\begin{eqnarray}
&&| Re(z)| \leq \sqrt{2 V_0},\nonumber\\
&&\sqrt{2 V_0-z^2}=\sqrt{\left(\sqrt{2 V_0}-z\right) \left(\sqrt{2 V_0}+z\right)}\nonumber\\
&&=r(z) e^{\frac{i}{2} (\rho (z)+\lambda (z)+2 \pi )}=-r(z) e^{\frac{i}{2} (\rho (z)+\lambda (z))};
\end{eqnarray}
\begin{eqnarray}
&&| Re(z)| >\sqrt{2 V_0},\nonumber\\
&&\sqrt{z^2-2 V_0}=\sqrt{\left(z-\sqrt{2 V_0}\right) \left(\sqrt{2 V_0}+z\right)}\nonumber\\
&&=r(z) e^{\frac{i}{2} (\rho (z)-\pi+\lambda (z) )}=-i r(z) e^{\frac{i}{2} (\rho (z)+\lambda (z))};
\end{eqnarray}
\[r(z)=\sqrt{\left| 2 V_0-z^2\right|},\rho (z)=\arg \left(\sqrt{2 V_0}-z\right),\]
\[\lambda (z)=\arg \left(\sqrt{2 V_0}+z\right).\]
Substituting Eqs. 88 and 89 into $J_1(k)$ and $J_2(k)$, we find that they are essentially the same function evaluated in different regions. We can thus combine the two integrals in Eq. 87 into one:
\begin{equation}
{f}_{j}(x)=Re\left(\int_{0}^{\infty }\frac{j \sqrt{2 \sigma }}{\pi ^2 j^2+2 \sigma ^2 V_0-\sigma ^2 k^2 } J(k)\,dk\right),
\end{equation}
\[J(k)=\frac{i k r(k) e^{\frac{i}{2}(\rho (k)+\lambda (k))} e^{r(k) (\sigma -x) e^{\frac{i}{2}(\rho (k)+\lambda (k))}}}{k^2-2 V_0}+\]
\[\frac{i k r(k) e^{\frac{i}{2}(\rho (k)+\lambda (k))} \left(i k r(k) e^{\frac{1}{2} i (\rho (k)+\lambda (k))}+k^2-2 V_0\right) }{V_0 \left(k^2-2 V_0\right) e^{-x r(k) e^{\frac{i}{2} (\rho (k)+\lambda (k))}}}\]
\[+\frac{i k r(k) \left(k \sqrt{k^2-2 V_0}-k^2+V_0\right) e^{\frac{i}{2}(\rho (k)+\lambda (k))} }{V_0 \left(k^2-2 V_0\right) e^{-r(k) (\sigma +x) e^{\frac{i}{2} (\rho (k)+\lambda (k))}}}.\]
For typesetting convenience, some numerator terms in the second and third terms of $J(k)$ have been moved below. Close inspection reveals that $J(k)$ is an even function on the real axis. Thus, the expansion function is:
\begin{equation}
{f}_{j}(x)=\frac{1}{2} Re\left(\int_{-\infty}^{+\infty }\frac{j \sqrt{2 \sigma }}{\pi ^2 j^2+2 \sigma ^2 V_0-\sigma ^2 k^2 } J(k)\,dk\right).
\end{equation}
Complexifying the integrand by substituting $k \to z$, we note that in the upper half-plane, $\frac{\pi}{2} < \frac{1}{2}\left(\lambda(z) + \rho(z)\right) < \frac{3\pi}{2}$, which satisfies the conditions of Jordan's lemma \cite{Jordan}. Integrating along the contour illustrated in Fig. 5c and applying the residue theorem \cite{Jordan}, we obtain the following result:
\begin{scriptsize}
\begin{eqnarray}
&&{f}_{j}(x)=\frac{1}{2} Re\left(\pi  i \left(\text{Res}\left(-\sqrt{2 V_0}\right)+\text{Res}\left(\sqrt{2 V_0}\right)\right)\right)+\nonumber\\
&&\frac{1}{2} Re\left(\pi  i \left(\text{Res}\left(-\sqrt{\frac{\pi ^2 j^2}{\sigma ^2}+2 V_0}\right)+\text{Res}\left(\sqrt{\frac{\pi ^2 j^2}{\sigma ^2}+2 V_0}\right)\right)\right)\nonumber\\
&&=\sqrt{\frac{2}{\sigma }} \sin \left(\frac{j \pi x}{\sigma }\right),0<x<\sigma .
\end{eqnarray}
\end{scriptsize}
The completeness proofs for the regions $x < 0$ and $x > \sigma$ are analogous to this and are omitted here to avoid redundancy. The proofs for the points $x = 0$ and $\sigma$ follow from the continuity of an improper integral depending on a parameter\cite{Zorich1}. Specifically, the integral in Eq. 87 at a fixed point $x$ is equal to the limit of Eq. 92 at the same fixed point $x$. Since the initial state is a continuous function, the completeness at the points $x = 0$ and $\sigma$ is thus established.

\paragraph{The initial state involves $x = 0$}
The initial state $\phi(x)$ in this case corresponds to $\tau < 0$ and $\tau + \sigma > 0$ in Eq. 1. Considering the region $-\chi < x < \chi$ where $\chi > \sigma$, we decompose $\phi(x)$ into two parts for $x \neq 0$: $\phi(x) = F(x) + G(x)$, where
\[F(x)=\begin{cases}
\phi (x), & -\chi <x<0\\ 
0, &0\leq x<\chi  
\end{cases};\]

\[G(x)=\begin{cases}
0, & -\chi <x\leq0\\ 
\phi (x), &0< x<\chi  
\end{cases}.\]
As established in the previous paragraph, $\left \{ \psi_\kappa(x)\right \}$ is complete with respect to the following two types of initial states:
\[{\phi }_{i}(x)=\begin{cases}
\sqrt{\frac{2}{\chi }} \sin \left(\frac{i \pi(x+\chi)}{\chi }\right), & -\chi <x<0\\ 
 0,& \text{otherwise}
\end{cases};\]

\[{\phi }_{j}(x)=\begin{cases}
0, & \text{otherwise}\\ 
\sqrt{\frac{2}{\chi }} \sin \left(\frac{j \pi(x+\chi)}{\chi }\right),& 0<x<\chi 
\end{cases}.\]
\[i,j=1,2,3....\]
The measurement probability amplitudes for these two types of initial states all satisfy:
\begin{footnotesize} %  \footnotesize, \scriptsize, \tiny
\[\left\langle \psi _{\kappa }|\phi _n\right\rangle =\int_{-\infty }^{+\infty } \left(\psi _{\kappa }(x)\right){}^* \phi _n(x) \, dx=\int_{-\chi }^{+\chi } \left(\psi _{\kappa }(x)\right){}^* \phi _n(x) \, dx.\]
\end{footnotesize}
As established in Section IIA, $\left \{ \phi_i(x)\right \}$ is complete with respect to $F(x)$, and $\left \{\phi_j(x)\right \}$ is complete with respect to $G(x)$. For $x \in (-\chi, \chi)$,
\[F(x)=\sum _i  \left\langle \left.\phi _i\right|F\right\rangle\phi _i(x) =\sum _i \left\langle \left.\phi _i\right|F\right\rangle  \int  \psi _{\kappa }(x) \left\langle \psi _{\kappa }|\phi _i\right\rangle  \, d\kappa\]
\[=\int\sum _i \left\langle \psi _{\kappa }|\phi _i\right\rangle  \left\langle \left.\phi _i\right|F\right\rangle   \psi _{\kappa }(x)\, d\kappa =\int  \left\langle \left.\psi _{\kappa }\right|F\right\rangle  \psi _{\kappa }(x) \, d\kappa,\]
\[G(x)=\int  \left\langle \left.\psi _{\kappa }\right|G\right\rangle  \psi _{\kappa }(x) \, d\kappa.\]
For $x\in(-\chi ,0)\cup(0,\chi )$,
\begin{eqnarray}
&&\phi(x) = F(x) + G(x)=\int  \left\langle \left.\psi _{\kappa }\right|F+G\right\rangle  \psi _{\kappa }(x) \, d\kappa \nonumber\\
&&=\int  \left\langle \left.\psi _{\kappa }\right|\phi \right\rangle  \psi _{\kappa }(x) \, d\kappa.
\end{eqnarray}

At $x = 0$, since the initial state $\phi(x)$ and eigenstates $\psi_\kappa(x)$ are continuous, it follows from the continuity of an integral depending on a parameter\cite{Zorich2} that:
\begin{equation}
    \int  \left\langle \left.\psi _{\kappa }\right|\phi (0)\right\rangle  \psi _{\kappa }(x) \, d\kappa=\lim_{x\to0}\phi (x)=\phi (0)
\end{equation}
It follows from Eqs. 93 and 94 that the eigenstate set $\left \{ \psi_\kappa(x)\right \}$ is complete with respect to $\phi(x)$ for $x \in (-\chi, \chi)$. However, since $\chi$ can be chosen arbitrarily large, in this sense, we conclude that $\left \{ \psi_\kappa(x)\right \}$ is complete with respect to $\phi(x)$ in the entire one-dimensional space.

\begin{figure}[t]  % t=顶部，b=底部，p=单独一页（PRL推荐的浮动体选项）
  \centering  % 图片居中
  % 关键：width=\linewidth 自动适配单栏宽度，keepaspectratio保持宽高比（避免拉伸）
  \includegraphics[width=\linewidth, keepaspectratio]{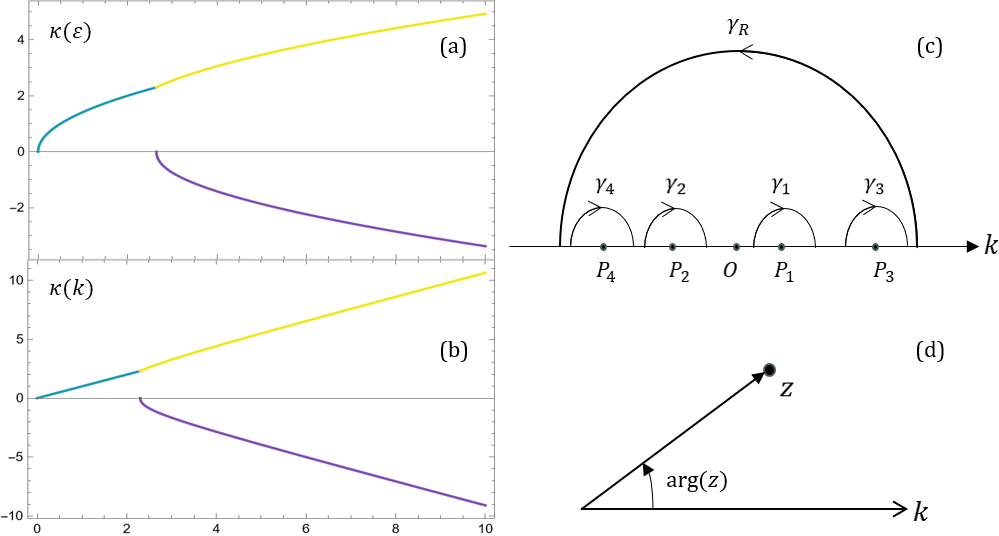}  % 图片文件（矢量图优先）
  \caption{Unequal constant potentiaL: (a)–(b)for $V_0=2.645$ band structure of the spectral function $\kappa$;(c)the integration path $\Gamma $;(d)the calculation method of argument range.}
  \label{fig:unequal constant potentia}
\end{figure}

\subsection{\label{sec:level22}Infinite potential on one side and constant potential on the other side}
In this case, common potentials include the one-sided open box potential and the exponential potential. The former has stronger physical relevance, while the latter has stronger mathematical relevance, and they are discussed separately below.

\subsubsection{\label{sec:level621}The set of eigenstates in one-sided open box potential}
We adopt natural units where \( \hbar = m = a =  1 \), the potential:
\begin{equation}
    V(x)=\begin{cases}
+\infty , & x<0\\ 
-V_0, &0<x<1 \\ 
0, & x>1
\end{cases}.
\end{equation}
where ${V}_{0}>0.$ From Eq. \ref{eq:SF} and 93, we find that its spectral function is ${\kappa}(\varepsilon)= \sqrt{2\varepsilon}$,energy eigenvalue$\varepsilon>0$.
This section shares some similarities with Section IIIB, so many of its aspects will be treated briefly.

\paragraph{Bound states}
For $-{V}_{0}<\varepsilon <0 $,
SE :
\[0<x<1,{\psi }''(x)+\beta ^{2}\psi(x) =0,\beta  =\sqrt{2(\varepsilon_n +{V}_{0})};\]
\[x>1,{\psi }''(x)-{{k} }^{2}\psi(x) =0,{k} =\sqrt{-2\varepsilon_n }.\]
The discrete energy levels $\varepsilon_n$ are determined by the energy level equation below.
\begin{eqnarray}
k=-\beta  \cot (\beta ).
\end{eqnarray}
From the continuity of ${\psi}$ and ${\psi}'$ at the boundary, we obtain the bound state equation as:
\begin{equation}
{\psi }_{n}(x)=A \begin{cases}
\sin (\beta x),& 0<x<1\\ 
e^{k} \sin \left(\beta \right) e^{-k x}, &x>1 
\end{cases}.
 \label{eq:eigenstates61}
\end{equation}
The positive real number $A$ is determined by normalization.

\paragraph{Free states} For $\varepsilon>0$,the SE:
\[0<x<1,{\psi }''(x)+\gamma ^{2}\psi(x) =0,\gamma =\sqrt{2(\varepsilon +{V}_{0})};\]
\[x>1,{\psi }''(x)+{\kappa}^{2}\psi(x) =0,{\kappa}=\sqrt{2\varepsilon }.\]
The free states are given by:
\begin{equation}
{\psi }_{\kappa}(x)=\begin{cases}
 \frac{\kappa \sin (\gamma x)}{\sqrt{\pi  \left(\gamma  ^2 \cos ^2{\gamma}+\kappa ^2 \sin ^2{\gamma}\right)/2}},& 0<x<1\\ 
\frac{ \kappa  \sin \gamma \cos (\kappa x-\kappa )+\gamma   \cos \gamma  \sin (\kappa x-\kappa)}{\sqrt{\pi  \left(\gamma  ^2 \cos ^2{\gamma}+\kappa ^2 \sin ^2{\gamma}\right)/2}}, & x>1
\end{cases}.
 \label{eq:eigenstates62}
\end{equation}
Where $\kappa(\varepsilon)$ is the spectral function. The solution of the above eigenstates relies on the method for calculating normalization coefficients of one-sided free states (Eq. 4).

\subsubsection{\label{sec:level622}The completeness of the set of eigenstates in one-sided open box potential}
Following Eqs. 18–22, we obtain the expansion functions for this potential; these functions are not presented here. We aim to verify the completeness of the set of eigenstates in this potential via numerical simulations. As shown in Fig. 6, with Fig. 6(a) as the original control case, the completeness is well maintained as the expansion functions vary with their parameters $\tau$ (Fig.6(b)), $\sigma$ (Fig. 6(c)), and $V_0$ (Fig. 6(d)).

For greater physical relevance, we enforce continuity of the potential inside and outside the box:
\[\mathcal{V}\left ( x\right )=\begin{cases}
+\infty, & x<0\\ 
-\mathcal{V}_{0}, & 0<x<2/3\\ 
-\mathcal{V}_{0}+3 \mathcal{V}_{0}(x-2/3), & 2/3<x<1\\ 
0, & x>1
\end{cases}.\]
As shown in Fig. 6(e)–(f), the completeness of the set of the eigenstates is still well maintained for the potential $\mathcal{V}(x)$.

For $V_0 = 0.893$, there are no discrete states (Fig. 6(c)), while for $V_0 (or  \mathcal{V}_{0}) = 2.645$, there is one discrete state. We also consider the case of $V_0 = 300.17$, which has eight discrete states. In this potential, we calculate the total measurement probability (a necessary condition for completeness to hold), and their deviations from 1 are within the order of $10^{-6}$, confirming that completeness is still well maintained.

\begin{figure}[t]  % t=顶部，b=底部，p=单独一页（PRL推荐的浮动体选项）
  \centering  % 图片居中
  % 关键：width=\linewidth 自动适配单栏宽度，keepaspectratio保持宽高比（避免拉伸）
  \includegraphics[width=\linewidth, keepaspectratio]{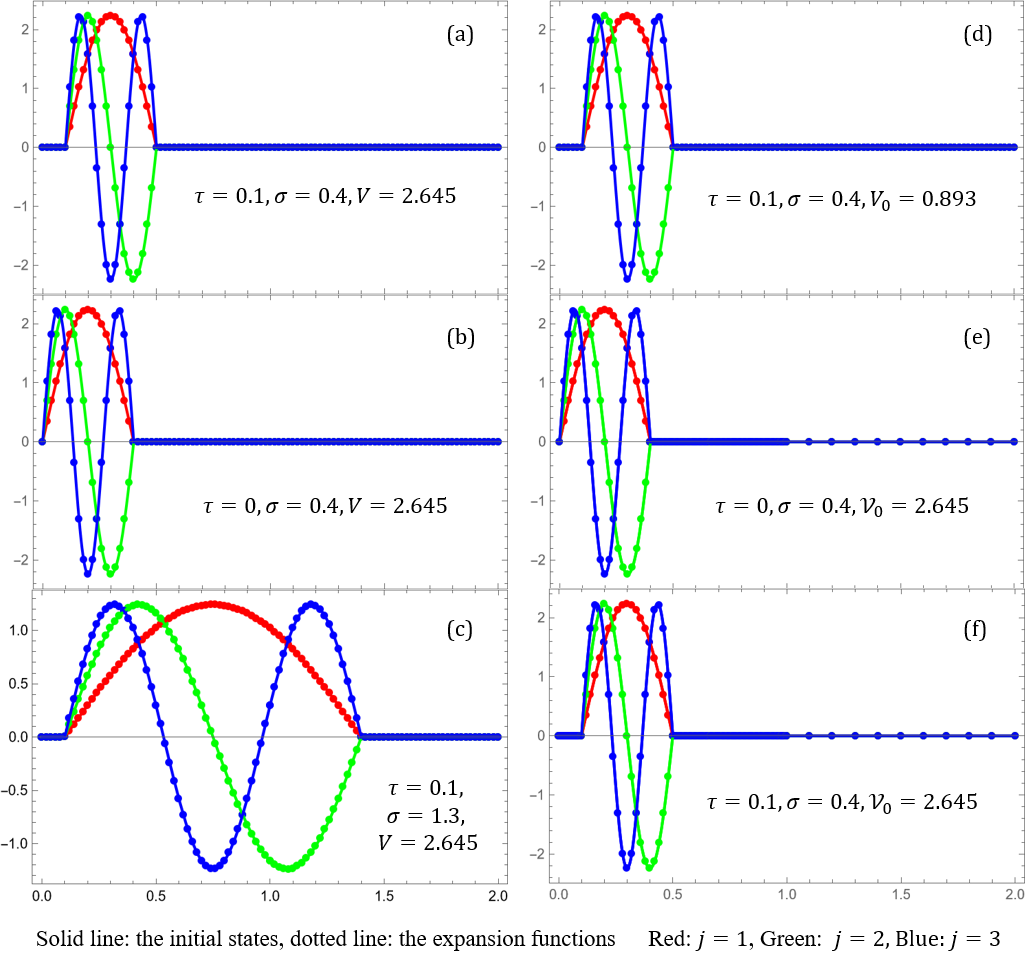}  % 图片文件（矢量图优先）
  \caption{(a)–(d) show comparisons between the expansion functions and the initial state for the one-sided open box potential $V(x)$, while (e)–(f) show those for the one-sided open box potential $\mathcal{V}(x)$.}
  \label{fig:one-sided open box potential}
\end{figure}

\subsection{\label{sec:level23}Infinite periodic potential on one side }
We first disregard the potential on the left and focus directly on the right side. The potential is given by:
\begin{equation}
   V(x)=\sum _{n=0 }^{+\infty } \gamma \delta  (x-n a),x>-b,0 < b < a.
\end{equation}
We adopt natural units where $\hbar = m = \gamma = 1$. 
The SE and eigenstate:
\begin{equation}\ x\ne 0,1,2...,\psi ''(x)+k^2 \psi (x)=0,k=\sqrt{2 \epsilon}.\end{equation}
\begin{equation}
    \psi (x)=A_n e^{i k x}+B_n e^{-i k x}.
\end{equation}
From the continuity of $\psi$ and the jump condition for the $\delta$-potential at $x = na$, we obtain:
\begin{equation}
    A_n e^{i a k n}+B_n e^{-i a k n}=A_{n+1} e^{i a k n}+B_{n+1} e^{-i a k n},
\end{equation}
\begin{eqnarray}
   && A_n \left(1+\frac{2 \gamma }{i k}\right) e^{i a k n}+B_n \left(-1+\frac{2 \gamma }{i k}\right) e^{-i a k n}\nonumber\\
   &&=A_{n+1} e^{i a k n}-B_{n+1} e^{-i a k n}.
\end{eqnarray}
Using linear algebra arguments, the general term formula is derived from the above recurrence relation as follows:
\begin{scriptsize} %  \footnotesize, \scriptsize, \tiny
\begin{eqnarray}
    {A_n}=\frac{ (-c-\lambda_2+1) \lambda_1^n+(c+\lambda_1-1) \lambda_2^n}{\lambda_1-\lambda_2}{A_0}-\frac{ c \left(\lambda_1^n-\lambda_2^n\right)}{\lambda_1-\lambda_1}B_0,
\end{eqnarray}
\begin{eqnarray}
    {B_n}=w^n \left(\frac{ c \left(\lambda_1^n-\lambda_2^n\right)}{w (\lambda_1-\lambda_2)}{A_0}+\frac{ (c+\lambda_1-1) \lambda_1^n+(-c-\lambda_2+1) \lambda_2^n}{\lambda_1-\lambda_2}{B_0}\right),
\end{eqnarray}
\begin{eqnarray*}
   && \lambda_1=\frac{\sqrt{v^2-4 w}+v}{2 w},\lambda_2=\frac{v-\sqrt{v^2-4 w}}{2 w},\nonumber\\
   &&v=(1-c) w+c+1,c=\frac{i \gamma }{k},w=e^{2 i a k}.
\end{eqnarray*}
\end{scriptsize}

\subsubsection{\label{sec:level71}The range of energy eigenvalues}
By the finiteness of the wavefunction $\lim_{x\to+\infty }\left | \psi (x)\right |$$=C$,  we obtain:
\begin{equation}
    \left | v\right |=\sqrt{\frac{{\gamma ^2 +k^2}}{k^2}}2\cos (k a-\alpha )\leq 2,
\end{equation}
\begin{equation}\left |\lambda _1\right |=\left |\lambda _1\right |=1,
\end{equation}
\begin{footnotesize} %  \footnotesize, \scriptsize, \tiny
\begin{eqnarray*}
&&v=\sqrt{\frac{{\gamma ^2 +k^2}}{k^2}}2\cos (k a-\alpha ){e}^{ik\alpha }=\left | v\right |{e}^{ik\alpha },\alpha =\tan^{-1}\frac{\gamma }{k}\nonumber\\
&&\lambda _1=\frac{\left | v\right | +i \sqrt{4-\left | v\right | ^2}}{2 e^{ika}},\lambda _2=\frac{\left | v\right | -i \sqrt{4-\left | v\right | ^2}}{2 e^{ika}}. 
\end{eqnarray*}
\end{footnotesize}
Performing trigonometric transformations on the finiteness equation (Eq. 106), we obtain:
\begin{equation}
    \cos (2\alpha )-\cos (2ka-2\alpha )\ge 0.
\end{equation}
By reducing the spectral equation (Eq. 43) in the Dirac comb, we obtain:
\begin{equation}
   -1\leq \cos ( \kappa a)=\frac{\gamma  \sin ( ka)}{k}+\cos (ka)\leq 1.
\end{equation}
Though differing in form, the two above inequalities can be unified into the same form when equality holds:
\begin{equation}
   \tan ( ka) (\tan (ka)-\tan (2 \alpha ))=0.
\end{equation}
In other words, the ranges of values for $k = \sqrt{2\varepsilon}$ (or equivalently for $\varepsilon$) are identical in both the one-sided and two-sided $\delta$-potentials. They delimit the same energy spectrum, as shown in Fig. 7a.

If the left side is another periodic $\delta$-potential, the energy spectrum is their overlapping interval, as shown in Fig. 7b. It is noteworthy that the above derivation imposes no constraints on the boundary conditions (i.e., the requirements on the values of $A_0$ and $B_0$).Equivalently, in the absence of an intermediate potential well, the energy spectrum is independent of the connecting potential. If an intermediate potential well is present, the energy spectrum acquires additional discrete values, with no modification to the continuous energy spectrum.

\subsubsection{\label{sec:level72}Orthonormalization of Eigenstates}
From the form of the eigenstates (Eq. 101), it can be seen that the eigenstates corresponding to different values of $k$ (or $\varepsilon$) are orthogonal. From Eq. 4, the normalization equation can be obtained:
\begin{equation}
    1=\underset{m\to +\infty }{\text{lim}}\frac{\int_0^{m a} \left(\frac{1}{\sqrt{\pi }}\right)^2 \, dx}{\sum _{n=1}^m \left(| A_n| ^2+| B_n| ^2\right) a}
\end{equation}
Through the simplification and solution of the aforementioned limit, we obtain:
\begin{equation}
    1=\frac{\sec \alpha  \sin (a k) \sin (a k-2 \alpha ) \csc (a k-\alpha )}{\pi \left (\mathcal{B} \sin (a k)-\mathcal{C} \cos (a k) \right ) }
\end{equation}
\[\mathcal{B}=|A_0| ^2+| B_0| ^2+i  (A_0 {B_0}^{*}-B_0 {A_0}^{*})\tan \alpha,\]
\[\mathcal{C}=| A_0+B_0| ^2 \tan \alpha.\]
If an infinitely high potential is introduced at $x = -b$, we obtain $B_0 = -e^{-2\mathrm{i} k b} A_0$. Without loss of generality, assuming that $A_0$ is a non-negative real number, we have:
\begin{equation}
    A_0=\sqrt{\frac{\sin (a k) \sin (a k-2 \alpha ) \csc (a k-\alpha )}{2 \pi  ( \cos (k (a-2 b) )\sin \alpha+\sin (a k-\alpha ))}}.
\end{equation}

To date, although both the eigenstates and the energy spectrum have been identified, the nature spectral function $\kappa(\varepsilon)$ for the expansion of eigenstates remains to be determined. Based on the content of Subsection IVC1, one might conjecture that $\kappa(\varepsilon)$ corresponds to the spectral function in the Dirac comb potential (Eq. 43). Unfortunately, upon verification, this is not the case. The spectral function for the one-sided $\delta-$potential requires further investigation.

\begin{figure}[t]  % t=顶部，b=底部，p=单独一页（PRL推荐的浮动体选项）
  \centering  % 图片居中
  % 关键：width=\linewidth 自动适配单栏宽度，keepaspectratio保持宽高比（避免拉伸）
  \includegraphics[width=\linewidth, keepaspectratio]{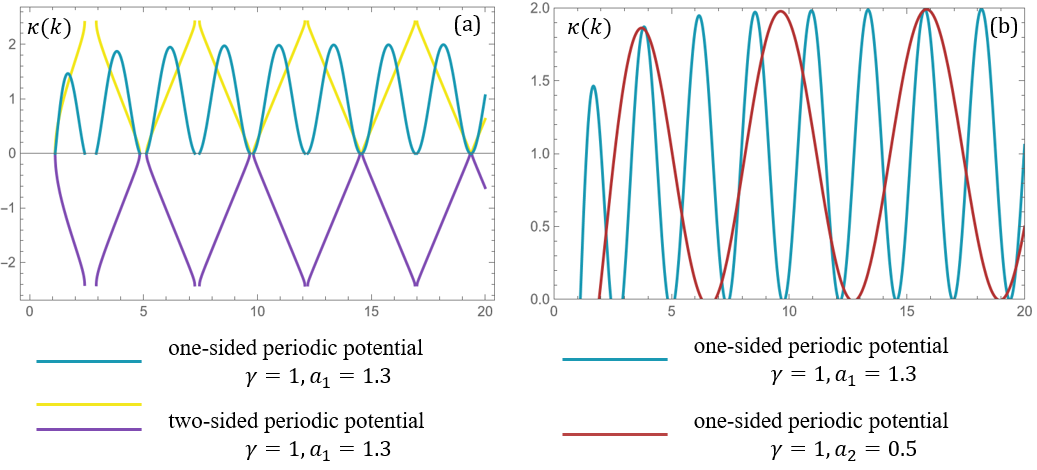}  % 图片文件（矢量图优先）
  \caption{$k = \sqrt{2\varepsilon}$,$\gamma$ is the strength of the $\delta$-potential, and $a$ is the potential period. (a) the one-sided periodic $\delta-$potential and the two-sided periodic $\delta-$potential have the same energy spectrum;(b) for the system with distinct periodic potentials on either side, the energy spectrum corresponds to the overlapping region of the energy spectra of the two periodic potentials.}
  \label{fig:Infinite periodic potential on one side}
\end{figure}

\subsection{\label{sec:level24}Finite periodic potential on one side }
In a one-sided finite periodic potential, the general energy eigenstates do not admit an elementary function form, making exact calculations---such as those performed in Subsection IVC1---difficult. Nevertheless, as can be found via numerical calculations, the range of the energy spectrum remains identical to that of the corresponding full periodic potential. If different periodic potentials are present on the left and right sides, the range of the energy spectrum is still the overlapping region of the energy spectra of the two periodic systems. This is consistent with the conclusion derived in Subsection IVC1, as shown in Fig.~7b.

From a physical perspective, there are five types of potentials in this section as follows. A one-sided infinite potential $+$ a one-sided finite periodic potential, e.g.:
\[V_1(x)=\begin{cases}
 +\infty ,&x<0 \\ 
\sin (2x), & x>0
\end{cases}.\]
A one-sided constant potential $+$ a one-sided finite periodic potential, e.g.:
\[V_2(x)=\begin{cases}
V_0 ,&x<0 \\ 
\sin (2x), & x>0
\end{cases}.\]
A one-sided finite periodic potential $+$ a one-sided finite periodic potential, e.g.:
\[V_3(x)=\sin (2x),-\infty <x<+\infty ;\]
\[V_4(x)=\begin{cases}
-\mathcal{V}_0\cos (\pi x/a), &x<0 \\ 
\mathcal{V}_1\cos (\pi x/b), & x>0
\end{cases}.\]
Two types of pseudoperiodic potentials:
\[V_4(x)=\cos \left(\frac{x}{| x| +1}\right),V_5(x)=\cos (x (| x| +1))\]
Except for $V_3(x)$, $\kappa(\varepsilon)$ remains intractable for the time being in other potentials (no relevant theories have been identified by the author). Translating $V_3(x)$ leftward by $\pi/4$ converts it to a cosine potential, with detailed solution methods provided in Section IIIC. We take $V_1(x)$ as an example to illustrate the challenges involved.

The SE:
\[\psi ''(x)+2 (\varepsilon -\sin (2 x)) \psi (x)=0,x>0.\]
Its general solution is given by:
\[\psi (x)=\mathcal{A} \text{MC}[2 \varepsilon,1,(x-\pi /4)]+\mathcal{B} \text{MS}[2 \varepsilon,1,(x-\pi /4)]\]
From the continuity of $\psi$ at $x=0$, we obtain
\[\mathcal{B}=-\frac{\text{MC}\left[2 \epsilon ,1,-\frac{\pi }{4}\right]}{\text{MS}\left[2 \epsilon ,1,-\frac{\pi }{4}\right]}\mathcal{A}.\]
% 最后一页关键段落前后添加
{\setlength{\parskip}{0.5\baselineskip} % 缩小段落间距（默认1\baselineskip）
Although we have determined the energy spectrum, the normalization equation---due to the fact that $\text{MC}$ and $\text{MS}$ do not admit an elementary function form---proves highly intractable to solve:
\begin{equation}
    \mathcal{A}^2={\lim_{a\rightarrow +\infty }\frac{\int_{0}^{a}{{{\left |{\frac{1}{\sqrt{\pi }}}\right |}^2}dx}}{\int_{0}^{a}{{{\left |\psi(x)\right |}^2}dx}}}.
\end{equation}
As for the spectral function $\kappa(\varepsilon)$ used for eigenstate expansion, its determination is even more challenging. Notably, while general mathematical methods for determining the energy spectrum of periodic potentials exist in the literature \cite{Arscott}\cite{Frank}, the determination of the energy spectrum for one-sided periodic potentials is established herein through our work.

\section{\label{sec:level3}Conclusions}
First, we delineated the scope of research on eigenstate completeness in quantum mechanics. Based on the limit of the potential function at infinity, the proof of completeness was divided into eight cases, for each of which theoretical proofs or numerical simulations are provided. 
Second, we presented the definition of orthonormalization for general free states (Eq. 2) and the solution to the normalization coefficients (Eqs. 3 and 4). 
Additionally, we proposed a general set of initial states (Eq. 1), which simplifies and concretizes the proof of completeness. 
Finally, we defined the spectral function for continuous energy eigenvalues (Eqs. 5, 27, 83, et al.). By taking the spectral function as the original integral variable of the expansion function, the relationship between the measured probability amplitude and the expansion function is endowed with the physical meaning of coordinate-momentum transformation (Eqs. 18–22, et al.).

Limitations and Future Directions: A key limitation is that we have not yet determined the spectral function in the one-sided periodic potential. Future work will focus on two aspects: unifying the eight types of potentials into a general theory, and extending the one-dimensional theory in this paper to the three-dimensional case.
} % 花括号限制作用域，不影响全局

\nocite{*}

\bibliography{ckwx}% Produces the bibliography via BibTeX.

\end{document}